\documentclass[reprint,amsmath,amssymb,aps,floatfix]{revtex4-1}

\usepackage{hyperref}

\usepackage{graphicx}
\usepackage{dcolumn}
\usepackage{bm}

\usepackage{physics}
\usepackage{siunitx}
\usepackage{mathtools}

\newcommand{\xzplane}{$x$-$z$ plane}

\newcommand{\ketdown}{\ket{\downarrow}}

\newcommand{\nm}{\text{~nm}}

\newcommand{\ns}{\text{~ns}}

\newcommand{\rad}{\text{ rad}}

\renewcommand{\phi}{\varphi}

\newcommand{\phiAC}{\phi_\text{}}

\newcommand{\udispl}{\abs{\alpha}}

\newcommand{\leffAC}{\lambda_\text{AC}}

\newcommand{\omegaLF}{\omega_\text{LF}}

\newcommand{\omegaQubit}{\omega_\text{S}}

\newcommand{\Deltatstrobo}{\Delta t}
\newcommand{\deltatstrobo}{\delta t}

\begin{document}
%
\title{Phase-Stable Traveling Waves Stroboscopically Matched for Super-Resolved Observation of Trapped-Ion Dynamics}

\author{Florian Hasse}
\author{Deviprasath Palani}
\author{Robin Thomm}
\author{Ulrich Warring}
\author{Tobias Schaetz}
\affiliation{University of Freiburg, Institut of Physics, Hermann-Herder-Strasse 3, Freiburg 79104, Germany}
\homepage{https://www.qsim.uni-freiburg.de/}

\date{\today}

\begin{abstract}
In quantum technologies, it is essential to understand and exploit the interplay of light and matter.
We introduce an approach, creating and maintaining the coherence of four oscillators: a global microwave reference field, a polarization-gradient traveling-wave pattern of light, and the spin and motional states of a single trapped ion.
The features of our method are showcased by probing the 140-nm periodic light pattern and stroboscopically tracing dynamical variations in position and momentum observables with noise floors of $1.8(2)\nm$ and $8(2)~$zN\,µs, respectively. 
The implications of our findings contribute to enhancing quantum control and metrological applications.
\end{abstract}

\maketitle

%
%
%
Quantum technologies are significantly enhancing control and sensing precision, e.g., by exploiting correlations, entanglement, and quantum non-demolition measurements\,\cite{Braginsky1980, Giovannetti2004, Degen2017}.
For instance, gravitational wave detection benefits from non-classical states of light and homo-/heterodyning measurement techniques\,\cite{Aasi2013, Acernese2019}.
Complementary, mechanical oscillators excel at sensing displacements caused by feeble forces\,\cite{Maiwald2009, Knuenz2010, Aspelmeyer2014, burd_quantum_2019, Gilmore2021}.
Trapped atomic ions have become fundamental in these explorations, offering insight into the details of underlying dynamics. 
Specifically, their well-isolated electronic/spin and motional/phonon states can be initialized, selectively coupled, and read out by high-fidelity operations\,\cite{leibfried_quantum_2003, wineland_nobel_2013}.
Central to these manipulations is the interplay of light and matter, with interactions that span a wide frequency spectrum, from mega- to petahertz, permitting innovations in quantum metrology, simulation, and computation\,\cite{Buluta2009, Ladd2010, wineland_nobel_2013}.
In this context, spatially structured light fields, with standing or traveling phase fronts, play an important role. 
They are realized via phase-coherent overlap of field sources, well controlled in frequency and polarization. 
Such alignments allow for setting up a diverse range of intensity and polarization-gradient wave patterns, applicable to atoms, ions, molecules, and nanoparticles\,\cite{metcalf_laser_1999}.
Choices between different setups, such as cavities or free space, depend on the specific task, as well as, its requirements on stability and controllability.
On the one hand, Fabry-Perot cavities provide highly stable phase patterns for spatial standing waves, yet, limiting dynamical control, constrained by cavity ring-down times, inhibiting fast modulations and traveling patterns\,\cite{Hood1998, Gutho2001, Mundt2002, Deist2022}.
On the other hand, free-space setups permit fast operations and control of dynamics, such as conditional displacement and squeezing via state-dependent optical dipole forces\,\cite{Leibfried2003, hume2011, Karpa2013, Miles2013, enderlein_single_2012, Schmiegelow2016, Affolter2020, Vasquez2023, Saner2023}.
Trapped ion systems have showcased standing wave patterns, as well as, \emph{walking} waves (in the MHz phonon-frame), using both active\,\cite{hume2011, Affolter2020, Saner2023} and passive\,\cite{Vasquez2023} stabilization methods.
But stability issues with \emph{running} wave patterns in the GHz spin-frame remain unaddressed or have been circumvented by laser-less couplings\,\cite{Ospelkaus2011, Srinivas2021}. 
While single-shot operations can be managed with transient phase coherence, maintaining stability across measurement series, or synchronizing multiple oscillators, demand (active) phase-stabilization techniques. 
In particular, running waves with high-bandwidth amplitude, frequency, and phase modulation can enable detailed explorations of the dynamics of spin-motional processes.

In this Letter, we actively phase synchronize a microwave (GHz) field with a 140-nm running wave, created by two counterpropagating UV-light (PHz) fields. 
When applied to a single trapped ion, this enables super-resolution spin-phase control, showcased in a hybrid Ramsey experiment. 
Using our setup's high modulation bandwidth, we stroboscopically apply the running wave to a coherently displaced ion, tracing its motional wave packet evolution in phase space (MHz). 
The established coherence links between our control fields and both, the spin and motional degrees, permit mapping positions and momenta with noise floors of $1.8(2)\nm$ and $8(2)~$zN\,µs, respectively.

\begin{figure}
\includegraphics{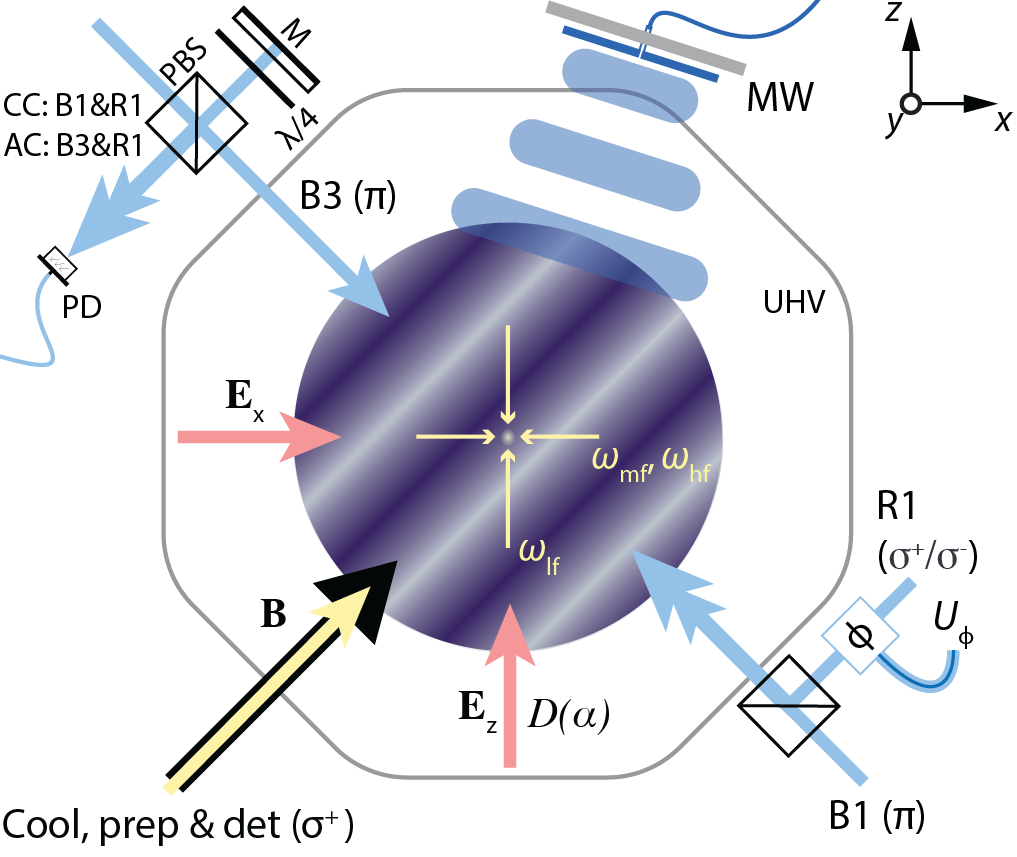}%
\caption{\label{fig1}
Experimental setup for synchronized, coherent control and analysis of the external (motional) and internal (electronic/spin) DOF of trapped ions (not to scale).
A single $^{25}$Mg$^+$ ion is spin-polarized and cooled close to the motional ground state while being confined by static and radio-frequency fields.
We exploit electric fields E$_{\text{x}}$, E$_{\text{z}}$ (red arrows), to statically offset or coherently displace the ion in the \xzplane.
We coherently control the spin-state by microwave fields (MW), while we exploit two-photon stimulated-Raman (TPSR) transitions, to render the spin-rotation sensitive to the motional states.
The corresponding laser-beams, B3 and R1, are aligned in antiparallel configuration (AC), forming a polarization-gradient traveling-wave pattern in the \xzplane. 
We synchronize the motional-oscillator and the two spin-oscillators (MW, AC) by locking the phase of the coherent displacement $D(\alpha)$ and the relative phase difference of TPSR beams to the phase of the MW. 
An interferometric setup for phase stabilization of the TPSR beams heterodyning the TPSR beams and mixing with the MW, allows deriving the required feedback signal $U_\Phi$.
}
\end{figure}

%
%
%
Our experimental setup, as illustrated in Fig.\,\ref{fig1} and detailed in our Supplement\,\cite{Note1}, integrates a linear radio-frequency ion trap with an operating frequency of \mbox{$\omega_{\text{RF}}/(2\pi) \simeq 56.3$~MHz} in a homogenous magnetic quantization field of \mbox{$|B| \simeq 0.58$~mT}.
%
%
A single $^{25}\text{Mg}^+$ ion (with nuclear spin $5/2$), selected for its suitable electronic $S_{1/2}$ hyperfine ground states, two of these states act as pseudo spin degree of freedom (DOF). 
The spin states are denoted as $\ket{\downarrow}$ and $\ket{\uparrow}$, with a transition frequency of $\omegaQubit/(2\pi) \simeq 1.8$~GHz. 
The three phonon DOF are described by three decoupled harmonic oscillators with $\omega_{\text{LF}}/(2\pi) \simeq 1.3$~MHz, the motion approximately along $z$ (axial), and $\omega_{\text{MF}}/(2\pi) \simeq 2.9$~MHz and $\omega_{\text{HF}}/(2\pi) \simeq 4.5$~MHz, motions in the $x$-$y$ (radial) plane at an angle of $\simeq30^{\circ}$ regarding the $y$ axis.
Cooling and detection lasers are tuned near a cycling transition between $S_{1/2}$ and $P_{3/2}$ substates and are aligned to the magnetic field, illustrated by a yellow arrow in Fig.\,\ref{fig1}. 
For repumping and $\ket{\downarrow}$-state preparation, we utilize beams, coupling appropriate Zeeman substates of $S_{1/2}$ and $P_{1/2}$. 
Fluorescence detection via a photomultiplier tube and subsequent photon histogram analysis, enables us to determine the electronic state population $P_{\downarrow}$ and we reconstruct population distributions of the motional states by mapping them onto the electronic states for calibration purposes\,\cite{leibfried_quantum_2003}.
Further, we use a camera to resolve ion positions within $\pm 1~$µm in the $x$-$z$ plane.
All normal modes of the ion are cooled near the three-dimensional ground state with low mean thermal occupation numbers $\langle n \rangle_{\text{th}} < 0.2$.
The effective size of the ion, i.e., the width of the ground state wave functions, is $\simeq10\nm$ in all three dimensions.
An arbitrary waveform generator is employed to initialize ions near $(\Delta x,\Delta y, \Delta z) = (\Delta x_0,\Delta y_0, \Delta z_0)= (0,0,0)$, to reposition the ion (static displacements), and to resonantly excite motional DOF (dynamic displacements):
For static displacements, pre-calibrated voltage sets applied to six control electrodes, shift the ion in the $x$-$z$ plane. 
Field amplitudes are kept between $\Delta E_{\text{x}} = \pm48.2$~V/m and $\Delta E_{\text{z}} = \pm7.2$~V/m, corresponding to static displacements of $<\pm200$~nm.
For dynamic displacements along the axial motion, we apply a coherent excitation pulse, $D(\alpha)$, via a single electrode, generating an oscillating electric field (red arrow in Fig.\,\ref{fig1}) pointing predominantly along $z$ at $\omega_{\text{LF}}$ for fixed durations, with tunable phases and amplitudes to initialize coherent displaced states with $\alpha = |\alpha| \exp(i\vartheta_0)$ -- realizing a displacement operation -- with tunable amplitude $|\alpha|$ and phase $\vartheta_0$.
We send a microwave (MW) reference signal via a biquad antenna to the ion to control/synchronize the spin DOF. 
In addition, we manipulate the electronic and motional DOF coherently using two-photon stimulated-Raman (TPSR) transitions.
We employ a UV laser system with 280~nm, detuned from the $S_{1/2}$-to-$P_{3/2}$ transition by $\Delta_{\text{R}}/(2\pi)\simeq20~$GHz, to apply TPSR couplings. 
The laser output beam is split multiple times via acousto-optic modulators (AOM) into three individually controllable beams: B1 ($\pi$-polarized), B3 ($\pi$-polarized), and R1 ($\sigma^+$/$\sigma^-$-polarized). 
The frequency difference between these beams can be fine-tuned around $\omegaQubit = \omega_{\text{B1}} - \omega_{\text{R1}} = \omega_{\text{B3}} - \omega_{\text{R1}}$.
Laser pulse (flash) durations are set minimally to $\deltatstrobo\simeq 100~$ns, currently limited by the speed of sound and laser beam waists in the AOMs.
The collimated beams are focused using 150-mm focal length lenses, yielding beam waists of $\simeq50~$µm near the ion with a Rayleigh length of $\simeq 30~$mm.
We utilize two distinct TPSR beam combinations: collinear configuration (CC with B1 and R1), with motional insensitive couplings, and antiparallel configuration (AC with B3 and R1), with motional sensitive couplings to all three normal modes. 
The effective AC wave vector $k_{\text{AC}}$ points along B3 towards R1, yielding a polarization-gradient traveling-wave pattern with a period of $\simeq140$~nm, and is used, e.g., for sideband cooling.
We establish and maintain phase coherence between AC, MW, and $D(\alpha)$ pulses via an active phase stabilization setup\,\cite{Note1}. 
The optical components of this system are housed within a solid aluminum block, approx. 20 cm squared, and individual laser beams B1, B3, and R1 are directed free-space through tubes over multiple meters, reducing disturbances from convection. 
The core components of the stabilization setup include a polarizing beam splitter (PBS), a quarter-wave plate ($\lambda/4$), a mirror (M), and a photodiode (PD) with a GHz bandwidth, as depicted in Fig.\,\ref{fig1}. 
When set to the phase-locked mode, the CC or AC beams generate heterodyne signals (GHz) on the PD at $\omegaQubit$, and the effective TPSR phase is referenced via a homodyne mixing of this laser beat note and the MW local oscillator. 
A control signal, $U_{\Phi}$, tunes a phase shifter, controlling the phase $\Phi$ of the signal, that is driving our AOMs. 
In this way, we establish an active tuning range of $\simeq 10.4~$rad at a bandwidth of $2\pi\,20~$kHz.
The coherence of the $D(\alpha)$ pulse with the MW signal is ensured by a 10-MHz GPS-referenced master clock that synchronizes (phase references) all relevant components of our classical control and data acquisition system.
Typical experimental sequences commence with an initialization section: 
Ion positions are set, and an optional phase \emph{lock-up} pulse for the TPSR combinations with a duration of $100~$µs is applied, followed by the preparation of motional and electronic states, lasting a few milliseconds.
Each sequence culminates with the detection of $30~$µs, and we repeat full sequences a few hundred times at fixed parameter settings.

\begin{figure}
\includegraphics{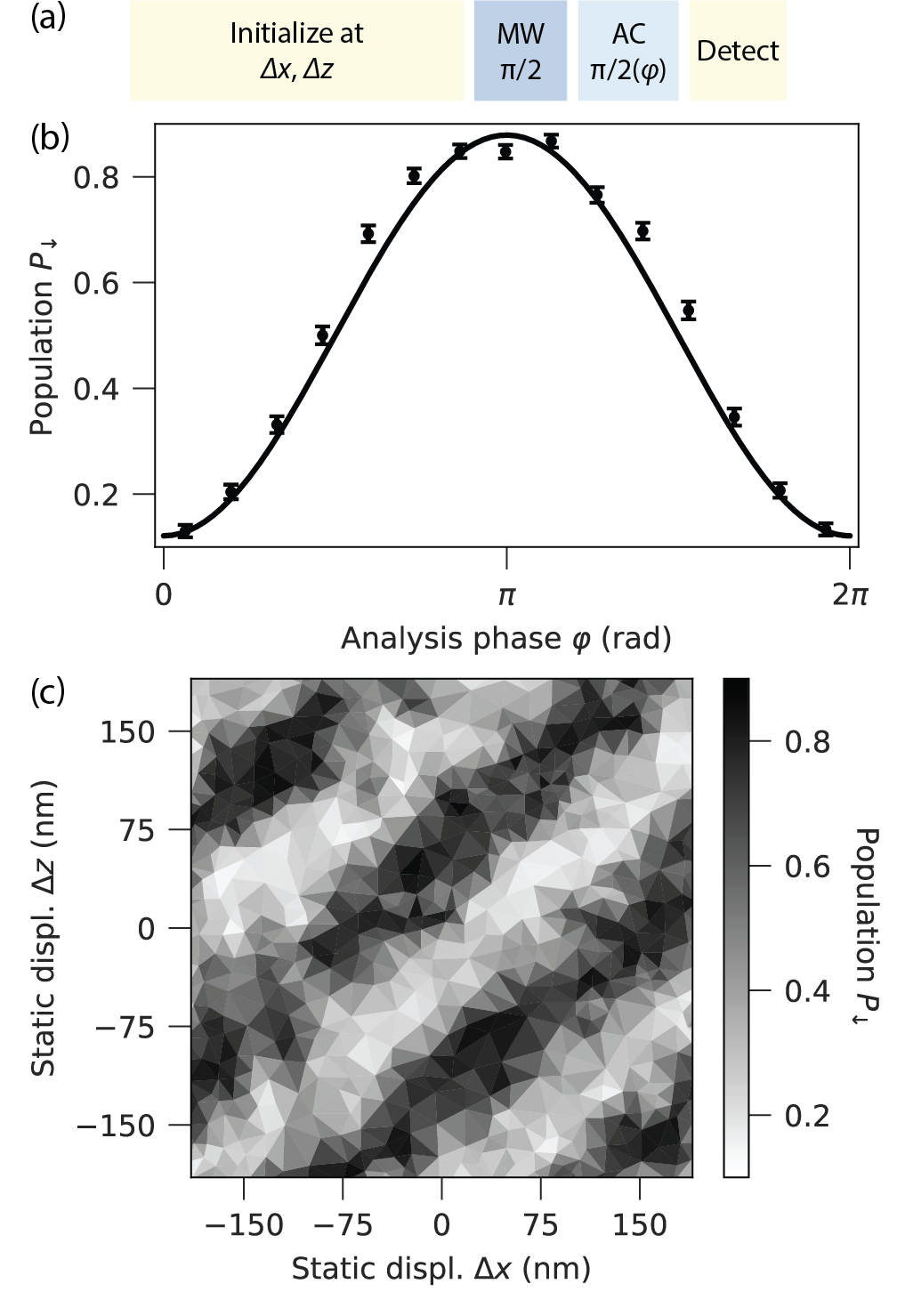}
\caption{\label{fig2} Probing the phase stability of the polarization-gradient traveling-wave pattern.
(a)~After initialization, the spin phase is synchronized using a global MW $\pi/2$ pulse. 
An AC pattern providing an analysis $\pi/2$ pulse with variable phase $\phi$, followed by state detection of $\ket{\downarrow}$.
(b)~With active-phase stabilization (data points), we achieve a coherence contrast of $0.76(3)$, constrained by AC beam path jitters of $\simeq10\nm$. 
The effects of both temperature and size on motion remain negligible.  
(c)~We reconstruct the two-dimensional AC phase fronts in the $x$-$z$ plane, signifying full wavelength displacements along the effective wave vector.
}
\end{figure}

%
%
%
In an initial benchmark of our active phase stabilization, we adopt a Ramsey-type experiment, cf. Fig.\,\ref{fig2}(a), and establish the following hybrid calibration sequence:
After initialization in $\ket{\downarrow}$, a spin-superposition state is generated by an MW $\pi/2$ (synchronization) pulse, and coherence is assessed via CC or AC $\pi/2$ (analysis) pulses with variable-phase $\varphi$. 
As an example, we present raw data for an AC sequence in Fig.\,2(b) with a coherence contrast of $0.76(3)$.
%
%
%
We infer a short-term stability of $0.206(9)\rad$ at 2 s, a long-term stability of $0.38(2)\rad$ at 40 s, and $0.65(5)\rad$ at 200 s. 
We attribute these remaining shifts to beam path variations ranging from $5-15\nm$ on timescales of tens of seconds and longer, which curtails the AC coherence and deem motional effects from finite mode temperatures and sizes to be negligible\,\cite{Note1}.
For the CC combination, we record phase stabilities: $0.156(5)\rad$ at 2 s, $0.026(1)\rad$ at 40 s, and $0.016(2)\rad$ at 200 s -- suggesting that the stability of the AC combination is not limited by our stabilization system.
To account for residual drifts, we can interleave calibration sequences to effectively reduce phase variations to $\lesssim0.1\rad$ on all relevant timescales\,\cite{Note1}. 

In the following demonstrations, we apply extensions of the Ramsey hybrid sequences with MW and AC, enabling localized spin-phase probing and control.  
In an initial application, we execute the Ramsey sequence with a fixed relative analysis phase of $\phiAC \simeq \pi/2$ for variable $\Delta x$ and $\Delta z$, to examine the AC pattern. 
The corresponding data is presented in Fig.\,\ref{fig2}(c), demonstrating the stability of our active phase stabilization system.
From fitting a sinusoidal model, we obtain $\leffAC=138(1)\nm$ and a pattern rotation of $0.840(7)\rad$ regarding the $z$-axis, in agreement with estimates based on geometric considerations.

\begin{figure}
\includegraphics{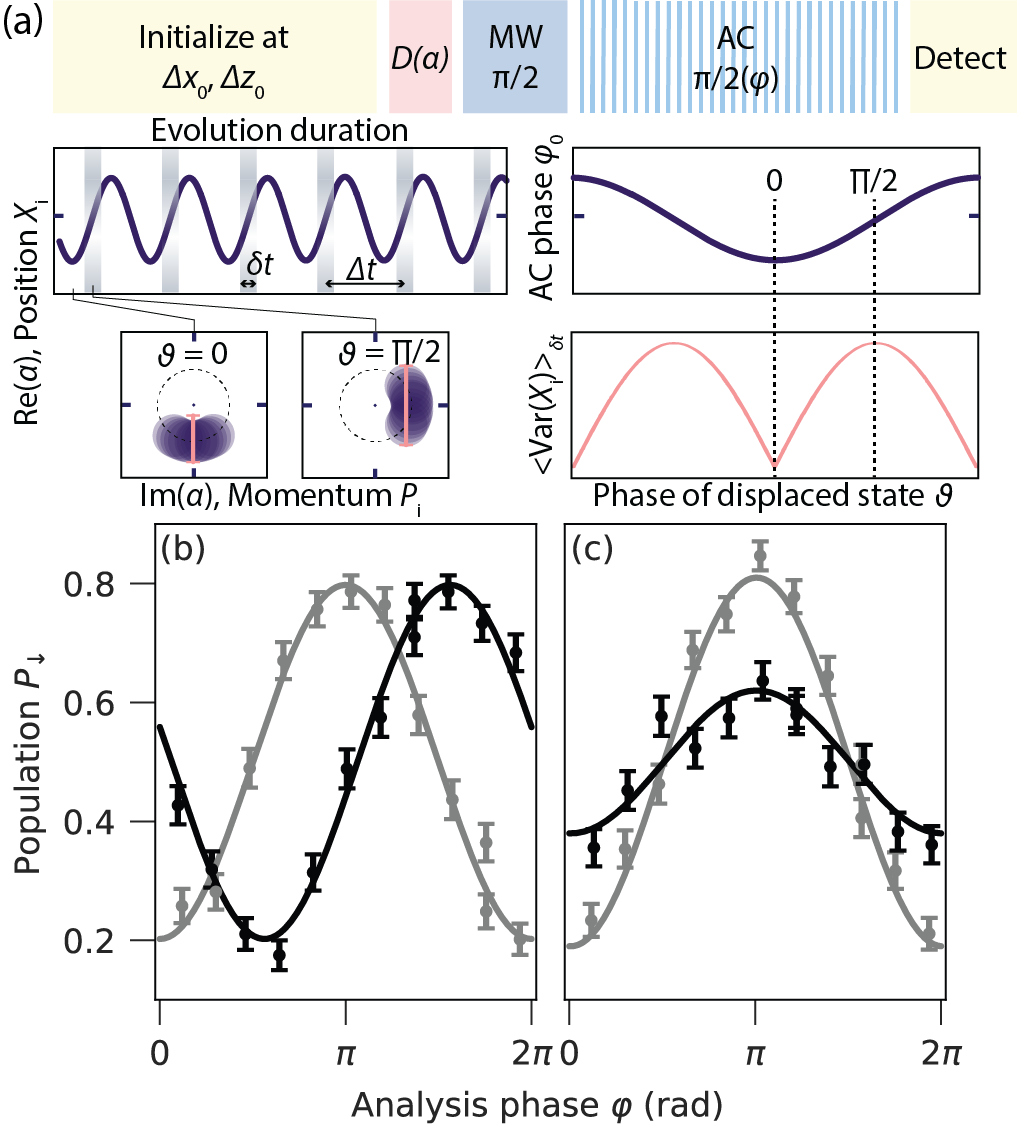}
\caption{\label{fig3}
Stroboscopic polarization-gradient traveling-wave pattern probing the dynamics of displaced states.
(a) We initialize a single ion and apply a coherent excitation pulse $D(\alpha)$ for a fixed duration, and variable phase $\vartheta_0$ and amplitude $\udispl$.
The sequence continues with a synchronization pulse, followed by a stroboscopic AC analysis pulse train of variable $\phi$, and concludes with $\ketdown$-state detection.
We illustrate the trajectory of a coherently displaced state and highlight tunable
pulse train parameters: $30$ flashes with duration $\deltatstrobo\simeq100\ns$, cycle duration $\Deltatstrobo=2\pi/\omegaLF$, and progressing phase to set up a complete $\pi/2$ pulse\,\cite{Note1}.
Phase-space diagram snapshots (not to scale) illustrate the time-averaged shapes of displaced states for $\vartheta=\{0, \pi/2\}$: The finite $\deltatstrobo$ yields a modulation of the time-averaged variance of the ion position $\langle\text{Var}(X_i)_{\deltatstrobo}\rangle$ as a function of $\vartheta=\vartheta_0+\omegaLF t$.
(b, c) We show experimental results (data points) with model fits (solid lines) as a function of $\phi$: For sets of $\udispl=\{0, 6.5(2)\}$ as gray, and black data points, respectively, and for $\vartheta_0 \simeq 0$ in (b) and $\vartheta_0 \simeq \pi/2$ in (c).
}
\end{figure}

\begin{figure}
\includegraphics{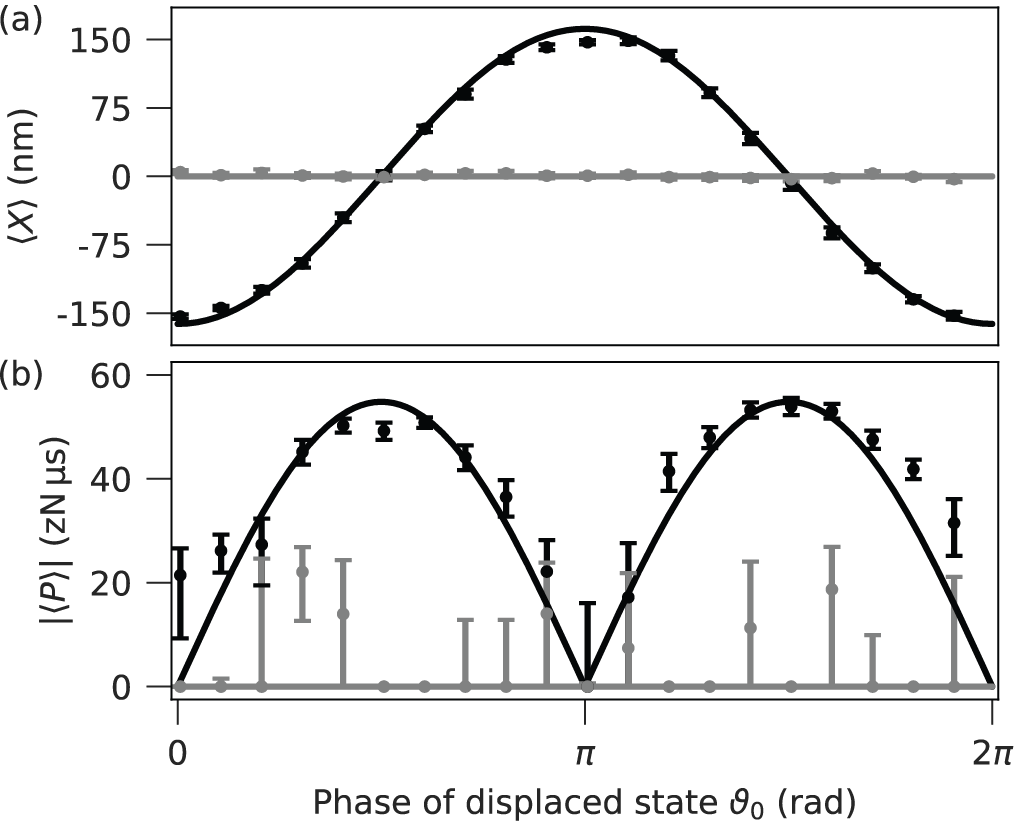}
\caption{\label{fig4}
Decoding of position and momentum of coherent displaced states.
(a) Position $\langle X \rangle$ and (b) the magnitude of momentum $|\langle P \rangle|$ are encoded in linear phase shifts and non-linear contrast variations in the spin DOF, respectively. 
We depict corresponding expectation values of coherent displaced states of the axial motion as a function of displacement phase $\vartheta_0$. Two discrete amplitudes are shown: $\udispl=0$ (gray) and 6.5(2) (black).
Solid lines represent ideal harmonic oscillator models, and we find noise floors of $1.8(2)\nm$ and $8(2)~$zN\,µs for position and momentum observables, respectively.
}
\end{figure}

In an advanced application, we resolve the dynamics of coherent displaced states of the axial mode, and we reconstruct expectation values of position $\langle X \rangle$ and momentum $|\langle P \rangle|$.
In Fig.\,\ref{fig3}(a), we illustrate the protocol, emphasizing the requirement of splitting the AC analysis pulse into a fine-tuned stroboscopic pulse train, $30$ flashes with flash durations $\delta t$\,\cite{Note1}. 
%
%
The cycling duration $\Deltatstrobo=2\pi/\omegaLF$ and other properties of the stroboscopic analysis pulse are matched to pre-calibrated displacement pulses $D(\alpha)$ that we apply before the synchronizing MW pulse.
We repeat this sequence for sets of $\vartheta_0$ and $|\alpha|$ (initializing variable displaced states) and probe $P_{\downarrow}$ as a function of $\phi$. 
Changes in position, $\langle X \rangle$, are linearly encoded in relative analysis phase shifts, $\phi_0$, while the magnitude of momenta, $|\langle P \rangle|$, are encoded non-linearly through contrast variations, $\Delta C$, due to finite time effects of the flashes\,\cite{Note1}. 
We illustrate the underlying principle in a series of time-averaged snapshots of phase-space, where the phase of the displaced state evolves $\vartheta = \vartheta_0 + \omegaLF\,t$, but yields smeared-out shapes increasing in size $\propto|\alpha|$: 
For $\vartheta \simeq 0$, momenta are lowest, the wave packet's effective size, given by the time-averaged variance of the ion position $\langle\text{Var}(X_i)_{\deltatstrobo}\rangle$, is minimal. That is, the phase of the running wave is well-defined, the contrast is highest, and $\phi_0 \propto |\alpha|$.
Conversely, for $\vartheta \simeq \pi/2$, momenta are largest, and the time-averaged wave packet size (in position space) is maximal. 
Consequently, the resulting averages over a larger distance, yield the lowest phase-scan contrast, minimal for the largest $|\alpha|$.
Corresponding samples of experimental results are shown in Fig.\,\ref{fig3}(b, c).
We yield dedicated decoding functions from numerical predictions for our chosen parameters and apply these to two data sets of variable $\vartheta_0$ and $|\alpha|=\{0, 6.5(2)\}$.
Figure\,\ref{fig4} shows decoded data together with coherent displaced state expectations that are calculated for fixed parameters\,\cite{Note1}.
%
%
We find qualitative agreement of data and idealized harmonic oscillator expectations. 
Current noise floors are $\simeq2\nm$ and $\simeq10~$zN\,µs for position and momentum observables, respectively. 
The sensitivity, dynamic range, and bandwidth of our sensing technique can be tuned by adjustments of the flash properties.
Numerically, we find: (i) that our stroboscopic method can probe, e.g., squeezed-state dynamics, and (ii) that the amount of back-action, change of motional state due to the probing, can be optimized for anticipated applications\,\cite{Note1}.
Additionally, by selectively adjusting modulation frequency and phase, our approach can be adapted to explore combinations of displaced and squeezed states, inferring individual amplitudes and phases of such superpositions. 
Integrating optical fibers/waveguides can enhance the stability of our method, and customized light modulation techniques can increase temporal control for resolving faster motional dynamics.

%
%
%
In conclusion, we actively stabilize a polarization-gradient traveling-wave pattern, probing the ion's position in sync with its motion. 
This demonstrates a coherent link between all our control fields and ensures prolonged phase stability beyond single-shot operations.
Our experimental results recorded via hybrid Ramsey sequences and calibrations guided by numerical studies enable us to condense the following key findings:
(i) We resolve the 140-nm light pattern with super-resolution using an effective probe size of approximately 10 nm, provided by the width of the ground-state wave function of a single atom.
(ii) We record variations in position and momentum observables of the dynamics of an excited state in phase space, here, exemplarily demonstrated for coherently displaced states in the MHz regime.
(iii) Our technique can be adapted to (cyclic) non-classical states, particularly squeezed states, to further increase sensitivities of position or momentum measurements.
(iv) And, notably, our modulated interactions can be fine-tuned either to enable minimally invasive measurements of phase-space properties or to allow for coherent, spin-phase dependent steering of phonon excitations.
In particular, we plan to expand our methods to transferring spatial entanglements, present in multimode-squeezed states\,\cite{Wittemer2019, Wittemer2020}, into the robust electronic degrees of freedom of multiple ions. 
Specifically, to continue the quest for studying analogs of relativistic quantum effects otherwise inaccessible to direct exploration, e.g., Hawking radiation and physics of the early universe. 
Generally, our work constitutes a contribution to advanced and versatile quantum control and measurement methods.

\begin{acknowledgments}
This work was supported by the Deutsche Forschungsgemeinschaft (DFG) (Grant No. SCHA 973/6-2) and the Georg~H.~Endress Foundation.
We acknowledge the fruitful discussions with J.~Bollinger and G.~Giedke. 
We thank F.~Grossmann and the workshop team for the fine mechanics work, J.~Denter for lab support, and F.~Thielemann for his graphics library \emph{minicoli}.
\end{acknowledgments}
%


\begin{thebibliography}{33}%
\makeatletter
\providecommand \@ifxundefined [1]{%
 \@ifx{#1\undefined}
}%
\providecommand \@ifnum [1]{%
 \ifnum #1\expandafter \@firstoftwo
 \else \expandafter \@secondoftwo
 \fi
}%
\providecommand \@ifx [1]{%
 \ifx #1\expandafter \@firstoftwo
 \else \expandafter \@secondoftwo
 \fi
}%
\providecommand \natexlab [1]{#1}%
\providecommand \enquote  [1]{``#1''}%
\providecommand \bibnamefont  [1]{#1}%
\providecommand \bibfnamefont [1]{#1}%
\providecommand \citenamefont [1]{#1}%
\providecommand \href@noop [0]{\@secondoftwo}%
\providecommand \href [0]{\begingroup \@sanitize@url \@href}%
\providecommand \@href[1]{\@@startlink{#1}\@@href}%
\providecommand \@@href[1]{\endgroup#1\@@endlink}%
\providecommand \@sanitize@url [0]{\catcode `\\12\catcode `\$12\catcode
  `\&12\catcode `\#12\catcode `\^12\catcode `\_12\catcode `\%12\relax}%
\providecommand \@@startlink[1]{}%
\providecommand \@@endlink[0]{}%
\providecommand \url  [0]{\begingroup\@sanitize@url \@url }%
\providecommand \@url [1]{\endgroup\@href {#1}{\urlprefix }}%
\providecommand \urlprefix  [0]{URL }%
\providecommand \Eprint [0]{\href }%
\providecommand \doibase [0]{http://dx.doi.org/}%
\providecommand \selectlanguage [0]{\@gobble}%
\providecommand \bibinfo  [0]{\@secondoftwo}%
\providecommand \bibfield  [0]{\@secondoftwo}%
\providecommand \translation [1]{[#1]}%
\providecommand \BibitemOpen [0]{}%
\providecommand \bibitemStop [0]{}%
\providecommand \bibitemNoStop [0]{.\EOS\space}%
\providecommand \EOS [0]{\spacefactor3000\relax}%
\providecommand \BibitemShut  [1]{\csname bibitem#1\endcsname}%
\let\auto@bib@innerbib\@empty
\bibitem [{\citenamefont {Braginsky}\ \emph {et~al.}(1980)\citenamefont
  {Braginsky}, \citenamefont {Vorontsov},\ and\ \citenamefont
  {Thorne}}]{Braginsky1980}%
  \BibitemOpen
  \bibfield  {author} {\bibinfo {author} {\bibfnamefont {V.~B.}\ \bibnamefont
  {Braginsky}}, \bibinfo {author} {\bibfnamefont {Y.~I.}\ \bibnamefont
  {Vorontsov}}, \ and\ \bibinfo {author} {\bibfnamefont {K.~S.}\ \bibnamefont
  {Thorne}},\ }\href {\doibase 10.1126/science.209.4456.547} {\bibfield
  {journal} {\bibinfo  {journal} {Science}\ }\textbf {\bibinfo {volume}
  {209}},\ \bibinfo {pages} {547} (\bibinfo {year} {1980})}\BibitemShut
  {NoStop}%
\bibitem [{\citenamefont {Giovannetti}\ \emph {et~al.}(2004)\citenamefont
  {Giovannetti}, \citenamefont {Lloyd},\ and\ \citenamefont
  {Maccone}}]{Giovannetti2004}%
  \BibitemOpen
  \bibfield  {author} {\bibinfo {author} {\bibfnamefont {V.}~\bibnamefont
  {Giovannetti}}, \bibinfo {author} {\bibfnamefont {S.}~\bibnamefont {Lloyd}},
  \ and\ \bibinfo {author} {\bibfnamefont {L.}~\bibnamefont {Maccone}},\ }\href
  {\doibase 10.1126/science.1104149} {\bibfield  {journal} {\bibinfo  {journal}
  {Science}\ }\textbf {\bibinfo {volume} {306}},\ \bibinfo {pages}
  {1330–1336} (\bibinfo {year} {2004})}\BibitemShut {NoStop}%
\bibitem [{\citenamefont {Degen}\ \emph {et~al.}(2017)\citenamefont {Degen},
  \citenamefont {Reinhard},\ and\ \citenamefont {Cappellaro}}]{Degen2017}%
  \BibitemOpen
  \bibfield  {author} {\bibinfo {author} {\bibfnamefont {C.~L.}\ \bibnamefont
  {Degen}}, \bibinfo {author} {\bibfnamefont {F.}~\bibnamefont {Reinhard}}, \
  and\ \bibinfo {author} {\bibfnamefont {P.}~\bibnamefont {Cappellaro}},\
  }\href {\doibase 10.1103/RevModPhys.89.035002} {\bibfield  {journal}
  {\bibinfo  {journal} {Rev. Mod. Phys.}\ }\textbf {\bibinfo {volume} {89}},\
  \bibinfo {pages} {035002} (\bibinfo {year} {2017})}\BibitemShut {NoStop}%
\bibitem [{\citenamefont {Aasi}\ \emph {et~al.}(2013)\citenamefont {Aasi} \emph
  {et~al.}}]{Aasi2013}%
  \BibitemOpen
  \bibfield  {author} {\bibinfo {author} {\bibfnamefont {J.}~\bibnamefont
  {Aasi}} \emph {et~al.},\ }\href {\doibase 10.1038/nphoton.2013.177}
  {\bibfield  {journal} {\bibinfo  {journal} {Nat. Photonics}\ }\textbf
  {\bibinfo {volume} {7}},\ \bibinfo {pages} {613} (\bibinfo {year}
  {2013})}\BibitemShut {NoStop}%
\bibitem [{\citenamefont {Acernese}\ \emph {et~al.}(2019)\citenamefont
  {Acernese} \emph {et~al.}}]{Acernese2019}%
  \BibitemOpen
  \bibfield  {author} {\bibinfo {author} {\bibfnamefont {F.}~\bibnamefont
  {Acernese}} \emph {et~al.} (\bibinfo {collaboration} {Virgo Collaboration}),\
  }\href {\doibase 10.1103/PhysRevLett.123.231108} {\bibfield  {journal}
  {\bibinfo  {journal} {Phys. Rev. Lett.}\ }\textbf {\bibinfo {volume} {123}},\
  \bibinfo {pages} {231108} (\bibinfo {year} {2019})}\BibitemShut {NoStop}%
\bibitem [{\citenamefont {Maiwald}\ \emph {et~al.}(2009)\citenamefont
  {Maiwald}, \citenamefont {Leibfried}, \citenamefont {Britton}, \citenamefont
  {Bergquist}, \citenamefont {Leuchs},\ and\ \citenamefont
  {Wineland}}]{Maiwald2009}%
  \BibitemOpen
  \bibfield  {author} {\bibinfo {author} {\bibfnamefont {R.}~\bibnamefont
  {Maiwald}}, \bibinfo {author} {\bibfnamefont {D.}~\bibnamefont {Leibfried}},
  \bibinfo {author} {\bibfnamefont {J.}~\bibnamefont {Britton}}, \bibinfo
  {author} {\bibfnamefont {J.~C.}\ \bibnamefont {Bergquist}}, \bibinfo {author}
  {\bibfnamefont {G.}~\bibnamefont {Leuchs}}, \ and\ \bibinfo {author}
  {\bibfnamefont {D.~J.}\ \bibnamefont {Wineland}},\ }\href {\doibase
  10.1038/nphys1311} {\bibfield  {journal} {\bibinfo  {journal} {Nat Phys}\
  }\textbf {\bibinfo {volume} {5}},\ \bibinfo {pages} {551–554} (\bibinfo
  {year} {2009})}\BibitemShut {NoStop}%
\bibitem [{\citenamefont {Kn\"unz}\ \emph {et~al.}(2010)\citenamefont
  {Kn\"unz}, \citenamefont {Herrmann}, \citenamefont {Batteiger}, \citenamefont
  {Saathoff}, \citenamefont {H\"ansch}, \citenamefont {Vahala},\ and\
  \citenamefont {Udem}}]{Knuenz2010}%
  \BibitemOpen
  \bibfield  {author} {\bibinfo {author} {\bibfnamefont {S.}~\bibnamefont
  {Kn\"unz}}, \bibinfo {author} {\bibfnamefont {M.}~\bibnamefont {Herrmann}},
  \bibinfo {author} {\bibfnamefont {V.}~\bibnamefont {Batteiger}}, \bibinfo
  {author} {\bibfnamefont {G.}~\bibnamefont {Saathoff}}, \bibinfo {author}
  {\bibfnamefont {T.~W.}\ \bibnamefont {H\"ansch}}, \bibinfo {author}
  {\bibfnamefont {K.}~\bibnamefont {Vahala}}, \ and\ \bibinfo {author}
  {\bibfnamefont {T.}~\bibnamefont {Udem}},\ }\href {\doibase
  10.1103/PhysRevLett.105.013004} {\bibfield  {journal} {\bibinfo  {journal}
  {Phys. Rev. Lett.}\ }\textbf {\bibinfo {volume} {105}},\ \bibinfo {pages}
  {013004} (\bibinfo {year} {2010})}\BibitemShut {NoStop}%
\bibitem [{\citenamefont {Aspelmeyer}\ \emph {et~al.}(2014)\citenamefont
  {Aspelmeyer}, \citenamefont {Kippenberg},\ and\ \citenamefont
  {Marquardt}}]{Aspelmeyer2014}%
  \BibitemOpen
  \bibfield  {author} {\bibinfo {author} {\bibfnamefont {M.}~\bibnamefont
  {Aspelmeyer}}, \bibinfo {author} {\bibfnamefont {T.~J.}\ \bibnamefont
  {Kippenberg}}, \ and\ \bibinfo {author} {\bibfnamefont {F.}~\bibnamefont
  {Marquardt}},\ }\href {\doibase 10.1103/revmodphys.86.1391} {\bibfield
  {journal} {\bibinfo  {journal} {Rev. Mod. Phy.}\ }\textbf {\bibinfo {volume}
  {86}},\ \bibinfo {pages} {1391} (\bibinfo {year} {2014})}\BibitemShut
  {NoStop}%
\bibitem [{\citenamefont {Burd}\ \emph {et~al.}(2019)\citenamefont {Burd},
  \citenamefont {Srinivas}, \citenamefont {Bollinger}, \citenamefont {Wilson},
  \citenamefont {Wineland}, \citenamefont {Leibfried}, \citenamefont
  {Slichter},\ and\ \citenamefont {Allcock}}]{burd_quantum_2019}%
  \BibitemOpen
  \bibfield  {author} {\bibinfo {author} {\bibfnamefont {S.~C.}\ \bibnamefont
  {Burd}}, \bibinfo {author} {\bibfnamefont {R.}~\bibnamefont {Srinivas}},
  \bibinfo {author} {\bibfnamefont {J.~J.}\ \bibnamefont {Bollinger}}, \bibinfo
  {author} {\bibfnamefont {A.~C.}\ \bibnamefont {Wilson}}, \bibinfo {author}
  {\bibfnamefont {D.~J.}\ \bibnamefont {Wineland}}, \bibinfo {author}
  {\bibfnamefont {D.}~\bibnamefont {Leibfried}}, \bibinfo {author}
  {\bibfnamefont {D.~H.}\ \bibnamefont {Slichter}}, \ and\ \bibinfo {author}
  {\bibfnamefont {D.~T.~C.}\ \bibnamefont {Allcock}},\ }\href {\doibase
  10.1126/science.aaw2884} {\bibfield  {journal} {\bibinfo  {journal}
  {Science}\ }\textbf {\bibinfo {volume} {364}},\ \bibinfo {pages} {1163}
  (\bibinfo {year} {2019})}\BibitemShut {NoStop}%
\bibitem [{\citenamefont {Gilmore}\ \emph {et~al.}(2021)\citenamefont
  {Gilmore}, \citenamefont {Affolter}, \citenamefont {Lewis-Swan},
  \citenamefont {Barberena}, \citenamefont {Jordan}, \citenamefont {Rey},\ and\
  \citenamefont {Bollinger}}]{Gilmore2021}%
  \BibitemOpen
  \bibfield  {author} {\bibinfo {author} {\bibfnamefont {K.~A.}\ \bibnamefont
  {Gilmore}}, \bibinfo {author} {\bibfnamefont {M.}~\bibnamefont {Affolter}},
  \bibinfo {author} {\bibfnamefont {R.~J.}\ \bibnamefont {Lewis-Swan}},
  \bibinfo {author} {\bibfnamefont {D.}~\bibnamefont {Barberena}}, \bibinfo
  {author} {\bibfnamefont {E.}~\bibnamefont {Jordan}}, \bibinfo {author}
  {\bibfnamefont {A.~M.}\ \bibnamefont {Rey}}, \ and\ \bibinfo {author}
  {\bibfnamefont {J.~J.}\ \bibnamefont {Bollinger}},\ }\href {\doibase
  10.1126/science.abi5226} {\bibfield  {journal} {\bibinfo  {journal}
  {Science}\ }\textbf {\bibinfo {volume} {373}},\ \bibinfo {pages} {673}
  (\bibinfo {year} {2021})}\BibitemShut {NoStop}%
\bibitem [{\citenamefont {Leibfried}\ \emph
  {et~al.}(2003{\natexlab{a}})\citenamefont {Leibfried}, \citenamefont {Blatt},
  \citenamefont {Monroe},\ and\ \citenamefont
  {Wineland}}]{leibfried_quantum_2003}%
  \BibitemOpen
  \bibfield  {author} {\bibinfo {author} {\bibfnamefont {D.}~\bibnamefont
  {Leibfried}}, \bibinfo {author} {\bibfnamefont {R.}~\bibnamefont {Blatt}},
  \bibinfo {author} {\bibfnamefont {C.}~\bibnamefont {Monroe}}, \ and\ \bibinfo
  {author} {\bibfnamefont {D.}~\bibnamefont {Wineland}},\ }\href {\doibase
  10.1103/RevModPhys.75.281} {\bibfield  {journal} {\bibinfo  {journal} {Rev.
  Mod. Phys.}\ }\textbf {\bibinfo {volume} {75}},\ \bibinfo {pages} {281}
  (\bibinfo {year} {2003}{\natexlab{a}})}\BibitemShut {NoStop}%
\bibitem [{\citenamefont {Wineland}(2013)}]{wineland_nobel_2013}%
  \BibitemOpen
  \bibfield  {author} {\bibinfo {author} {\bibfnamefont {D.~J.}\ \bibnamefont
  {Wineland}},\ }\href {\doibase 10.1103/RevModPhys.85.1103} {\bibfield
  {journal} {\bibinfo  {journal} {Rev. Mod. Phys.}\ }\textbf {\bibinfo {volume}
  {85}},\ \bibinfo {pages} {1103} (\bibinfo {year} {2013})}\BibitemShut
  {NoStop}%
\bibitem [{\citenamefont {Buluta}\ and\ \citenamefont
  {Nori}(2009)}]{Buluta2009}%
  \BibitemOpen
  \bibfield  {author} {\bibinfo {author} {\bibfnamefont {I.}~\bibnamefont
  {Buluta}}\ and\ \bibinfo {author} {\bibfnamefont {F.}~\bibnamefont {Nori}},\
  }\href {\doibase 10.1126/science.1177838} {\bibfield  {journal} {\bibinfo
  {journal} {Science}\ }\textbf {\bibinfo {volume} {326}},\ \bibinfo {pages}
  {108} (\bibinfo {year} {2009})}\BibitemShut {NoStop}%
\bibitem [{\citenamefont {Ladd}\ \emph {et~al.}(2010)\citenamefont {Ladd},
  \citenamefont {Jelezko}, \citenamefont {Laflamme}, \citenamefont {Nakamura},
  \citenamefont {Monroe},\ and\ \citenamefont {O'Brien}}]{Ladd2010}%
  \BibitemOpen
  \bibfield  {author} {\bibinfo {author} {\bibfnamefont {T.~D.}\ \bibnamefont
  {Ladd}}, \bibinfo {author} {\bibfnamefont {F.}~\bibnamefont {Jelezko}},
  \bibinfo {author} {\bibfnamefont {R.}~\bibnamefont {Laflamme}}, \bibinfo
  {author} {\bibfnamefont {Y.}~\bibnamefont {Nakamura}}, \bibinfo {author}
  {\bibfnamefont {C.}~\bibnamefont {Monroe}}, \ and\ \bibinfo {author}
  {\bibfnamefont {J.~L.}\ \bibnamefont {O'Brien}},\ }\href {\doibase
  10.1038/nature08812} {\bibfield  {journal} {\bibinfo  {journal} {Nature}\
  }\textbf {\bibinfo {volume} {464}},\ \bibinfo {pages} {45} (\bibinfo {year}
  {2010})}\BibitemShut {NoStop}%
\bibitem [{\citenamefont {Metcalf}\ and\ \citenamefont {Van
  Der~Straten}(1999)}]{metcalf_laser_1999}%
  \BibitemOpen
  \bibfield  {author} {\bibinfo {author} {\bibfnamefont {H.}~\bibnamefont
  {Metcalf}}\ and\ \bibinfo {author} {\bibfnamefont {P.}~\bibnamefont {Van
  Der~Straten}},\ }\href@noop {} {\emph {\bibinfo {title} {Laser {{Cooling}}
  and {{Trapping}}}}}\ (\bibinfo  {publisher} {{Springer}},\ \bibinfo {year}
  {1999})\BibitemShut {NoStop}%
\bibitem [{\citenamefont {Hood}\ \emph {et~al.}(1998)\citenamefont {Hood},
  \citenamefont {Chapman}, \citenamefont {Lynn},\ and\ \citenamefont
  {Kimble}}]{Hood1998}%
  \BibitemOpen
  \bibfield  {author} {\bibinfo {author} {\bibfnamefont {C.~J.}\ \bibnamefont
  {Hood}}, \bibinfo {author} {\bibfnamefont {M.~S.}\ \bibnamefont {Chapman}},
  \bibinfo {author} {\bibfnamefont {T.~W.}\ \bibnamefont {Lynn}}, \ and\
  \bibinfo {author} {\bibfnamefont {H.~J.}\ \bibnamefont {Kimble}},\ }\href
  {\doibase 10.1103/PhysRevLett.80.4157} {\bibfield  {journal} {\bibinfo
  {journal} {Phys. Rev. Lett.}\ }\textbf {\bibinfo {volume} {80}},\ \bibinfo
  {pages} {4157} (\bibinfo {year} {1998})}\BibitemShut {NoStop}%
\bibitem [{\citenamefont {Guth{\"o}hrlein}\ \emph {et~al.}(2001)\citenamefont
  {Guth{\"o}hrlein}, \citenamefont {Keller}, \citenamefont {Hayasaka},
  \citenamefont {Lange},\ and\ \citenamefont {Walther}}]{Gutho2001}%
  \BibitemOpen
  \bibfield  {author} {\bibinfo {author} {\bibfnamefont {G.~R.}\ \bibnamefont
  {Guth{\"o}hrlein}}, \bibinfo {author} {\bibfnamefont {M.}~\bibnamefont
  {Keller}}, \bibinfo {author} {\bibfnamefont {K.}~\bibnamefont {Hayasaka}},
  \bibinfo {author} {\bibfnamefont {W.}~\bibnamefont {Lange}}, \ and\ \bibinfo
  {author} {\bibfnamefont {H.}~\bibnamefont {Walther}},\ }\href {\doibase
  10.1038/35102129} {\bibfield  {journal} {\bibinfo  {journal} {Nature}\
  }\textbf {\bibinfo {volume} {414}},\ \bibinfo {pages} {49} (\bibinfo {year}
  {2001})}\BibitemShut {NoStop}%
\bibitem [{\citenamefont {Mundt}\ \emph {et~al.}(2002)\citenamefont {Mundt},
  \citenamefont {Kreuter}, \citenamefont {Becher}, \citenamefont {Leibfried},
  \citenamefont {Eschner}, \citenamefont {Schmidt-Kaler},\ and\ \citenamefont
  {Blatt}}]{Mundt2002}%
  \BibitemOpen
  \bibfield  {author} {\bibinfo {author} {\bibfnamefont {A.~B.}\ \bibnamefont
  {Mundt}}, \bibinfo {author} {\bibfnamefont {A.}~\bibnamefont {Kreuter}},
  \bibinfo {author} {\bibfnamefont {C.}~\bibnamefont {Becher}}, \bibinfo
  {author} {\bibfnamefont {D.}~\bibnamefont {Leibfried}}, \bibinfo {author}
  {\bibfnamefont {J.}~\bibnamefont {Eschner}}, \bibinfo {author} {\bibfnamefont
  {F.}~\bibnamefont {Schmidt-Kaler}}, \ and\ \bibinfo {author} {\bibfnamefont
  {R.}~\bibnamefont {Blatt}},\ }\href {\doibase 10.1103/PhysRevLett.89.103001}
  {\bibfield  {journal} {\bibinfo  {journal} {Phys. Rev. Lett.}\ }\textbf
  {\bibinfo {volume} {89}},\ \bibinfo {pages} {103001} (\bibinfo {year}
  {2002})}\BibitemShut {NoStop}%
\bibitem [{\citenamefont {Deist}\ \emph {et~al.}(2022)\citenamefont {Deist},
  \citenamefont {Gerber}, \citenamefont {Lu}, \citenamefont {Zeiher},\ and\
  \citenamefont {Stamper-Kurn}}]{Deist2022}%
  \BibitemOpen
  \bibfield  {author} {\bibinfo {author} {\bibfnamefont {E.}~\bibnamefont
  {Deist}}, \bibinfo {author} {\bibfnamefont {J.~A.}\ \bibnamefont {Gerber}},
  \bibinfo {author} {\bibfnamefont {Y.-H.}\ \bibnamefont {Lu}}, \bibinfo
  {author} {\bibfnamefont {J.}~\bibnamefont {Zeiher}}, \ and\ \bibinfo {author}
  {\bibfnamefont {D.~M.}\ \bibnamefont {Stamper-Kurn}},\ }\href {\doibase
  10.1103/PhysRevLett.128.083201} {\bibfield  {journal} {\bibinfo  {journal}
  {Phys. Rev. Lett.}\ }\textbf {\bibinfo {volume} {128}},\ \bibinfo {pages}
  {083201} (\bibinfo {year} {2022})}\BibitemShut {NoStop}%
\bibitem [{\citenamefont {Leibfried}\ \emph
  {et~al.}(2003{\natexlab{b}})\citenamefont {Leibfried}, \citenamefont
  {DeMarco}, \citenamefont {Meyer}, \citenamefont {Lucas}, \citenamefont
  {Barrett}, \citenamefont {Britton}, \citenamefont {Itano}, \citenamefont
  {Jelenkovi{\'c}}, \citenamefont {Langer}, \citenamefont {Rosenband},\ and\
  \citenamefont {Wineland}}]{Leibfried2003}%
  \BibitemOpen
  \bibfield  {author} {\bibinfo {author} {\bibfnamefont {D.}~\bibnamefont
  {Leibfried}}, \bibinfo {author} {\bibfnamefont {B.}~\bibnamefont {DeMarco}},
  \bibinfo {author} {\bibfnamefont {V.}~\bibnamefont {Meyer}}, \bibinfo
  {author} {\bibfnamefont {D.}~\bibnamefont {Lucas}}, \bibinfo {author}
  {\bibfnamefont {M.}~\bibnamefont {Barrett}}, \bibinfo {author} {\bibfnamefont
  {J.}~\bibnamefont {Britton}}, \bibinfo {author} {\bibfnamefont {W.~M.}\
  \bibnamefont {Itano}}, \bibinfo {author} {\bibfnamefont {B.}~\bibnamefont
  {Jelenkovi{\'c}}}, \bibinfo {author} {\bibfnamefont {C.}~\bibnamefont
  {Langer}}, \bibinfo {author} {\bibfnamefont {T.}~\bibnamefont {Rosenband}}, \
  and\ \bibinfo {author} {\bibfnamefont {D.~J.}\ \bibnamefont {Wineland}},\
  }\href {\doibase 10.1038/nature01492} {\bibfield  {journal} {\bibinfo
  {journal} {Nature}\ }\textbf {\bibinfo {volume} {422}},\ \bibinfo {pages}
  {412} (\bibinfo {year} {2003}{\natexlab{b}})}\BibitemShut {NoStop}%
\bibitem [{\citenamefont {Hume}\ \emph {et~al.}(2011)\citenamefont {Hume},
  \citenamefont {Chou}, \citenamefont {Leibrandt}, \citenamefont {Thorpe},
  \citenamefont {Wineland},\ and\ \citenamefont {Rosenband}}]{hume2011}%
  \BibitemOpen
  \bibfield  {author} {\bibinfo {author} {\bibfnamefont {D.~B.}\ \bibnamefont
  {Hume}}, \bibinfo {author} {\bibfnamefont {C.~W.}\ \bibnamefont {Chou}},
  \bibinfo {author} {\bibfnamefont {D.~R.}\ \bibnamefont {Leibrandt}}, \bibinfo
  {author} {\bibfnamefont {M.~J.}\ \bibnamefont {Thorpe}}, \bibinfo {author}
  {\bibfnamefont {D.~J.}\ \bibnamefont {Wineland}}, \ and\ \bibinfo {author}
  {\bibfnamefont {T.}~\bibnamefont {Rosenband}},\ }\href {\doibase
  10.1103/PhysRevLett.107.243902} {\bibfield  {journal} {\bibinfo  {journal}
  {Phys. Rev. Lett.}\ }\textbf {\bibinfo {volume} {107}},\ \bibinfo {pages}
  {243902} (\bibinfo {year} {2011})}\BibitemShut {NoStop}%
\bibitem [{\citenamefont {Karpa}\ \emph {et~al.}(2013)\citenamefont {Karpa},
  \citenamefont {Bylinskii}, \citenamefont {Gangloff}, \citenamefont {Cetina},\
  and\ \citenamefont {Vuleti\ifmmode~\acute{c}\else \'{c}\fi{}}}]{Karpa2013}%
  \BibitemOpen
  \bibfield  {author} {\bibinfo {author} {\bibfnamefont {L.}~\bibnamefont
  {Karpa}}, \bibinfo {author} {\bibfnamefont {A.}~\bibnamefont {Bylinskii}},
  \bibinfo {author} {\bibfnamefont {D.}~\bibnamefont {Gangloff}}, \bibinfo
  {author} {\bibfnamefont {M.}~\bibnamefont {Cetina}}, \ and\ \bibinfo {author}
  {\bibfnamefont {V.}~\bibnamefont {Vuleti\ifmmode~\acute{c}\else
  \'{c}\fi{}}},\ }\href {\doibase 10.1103/PhysRevLett.111.163002} {\bibfield
  {journal} {\bibinfo  {journal} {Phys. Rev. Lett.}\ }\textbf {\bibinfo
  {volume} {111}},\ \bibinfo {pages} {163002} (\bibinfo {year}
  {2013})}\BibitemShut {NoStop}%
\bibitem [{\citenamefont {Miles}\ \emph {et~al.}(2013)\citenamefont {Miles},
  \citenamefont {Simmons},\ and\ \citenamefont {Yavuz}}]{Miles2013}%
  \BibitemOpen
  \bibfield  {author} {\bibinfo {author} {\bibfnamefont {J.~A.}\ \bibnamefont
  {Miles}}, \bibinfo {author} {\bibfnamefont {Z.~J.}\ \bibnamefont {Simmons}},
  \ and\ \bibinfo {author} {\bibfnamefont {D.~D.}\ \bibnamefont {Yavuz}},\
  }\href {\doibase 10.1103/PhysRevX.3.031014} {\bibfield  {journal} {\bibinfo
  {journal} {Phys. Rev. X}\ }\textbf {\bibinfo {volume} {3}},\ \bibinfo {pages}
  {031014} (\bibinfo {year} {2013})}\BibitemShut {NoStop}%
\bibitem [{\citenamefont {Enderlein}\ \emph {et~al.}(2012)\citenamefont
  {Enderlein}, \citenamefont {Huber}, \citenamefont {Schneider},\ and\
  \citenamefont {Schaetz}}]{enderlein_single_2012}%
  \BibitemOpen
  \bibfield  {author} {\bibinfo {author} {\bibfnamefont {M.}~\bibnamefont
  {Enderlein}}, \bibinfo {author} {\bibfnamefont {T.}~\bibnamefont {Huber}},
  \bibinfo {author} {\bibfnamefont {C.}~\bibnamefont {Schneider}}, \ and\
  \bibinfo {author} {\bibfnamefont {T.}~\bibnamefont {Schaetz}},\ }\href
  {\doibase 10.1103/PhysRevLett.109.233004} {\bibfield  {journal} {\bibinfo
  {journal} {Phys. Rev. Lett.}\ }\textbf {\bibinfo {volume} {109}},\ \bibinfo
  {pages} {233004} (\bibinfo {year} {2012})}\BibitemShut {NoStop}%
\bibitem [{\citenamefont {Schmiegelow}\ \emph {et~al.}(2016)\citenamefont
  {Schmiegelow}, \citenamefont {Kaufmann}, \citenamefont {Ruster},
  \citenamefont {Schulz}, \citenamefont {Kaushal}, \citenamefont {Hettrich},
  \citenamefont {Schmidt-Kaler},\ and\ \citenamefont
  {Poschinger}}]{Schmiegelow2016}%
  \BibitemOpen
  \bibfield  {author} {\bibinfo {author} {\bibfnamefont {C.~T.}\ \bibnamefont
  {Schmiegelow}}, \bibinfo {author} {\bibfnamefont {H.}~\bibnamefont
  {Kaufmann}}, \bibinfo {author} {\bibfnamefont {T.}~\bibnamefont {Ruster}},
  \bibinfo {author} {\bibfnamefont {J.}~\bibnamefont {Schulz}}, \bibinfo
  {author} {\bibfnamefont {V.}~\bibnamefont {Kaushal}}, \bibinfo {author}
  {\bibfnamefont {M.}~\bibnamefont {Hettrich}}, \bibinfo {author}
  {\bibfnamefont {F.}~\bibnamefont {Schmidt-Kaler}}, \ and\ \bibinfo {author}
  {\bibfnamefont {U.~G.}\ \bibnamefont {Poschinger}},\ }\href {\doibase
  10.1103/PhysRevLett.116.033002} {\bibfield  {journal} {\bibinfo  {journal}
  {Phys. Rev. Lett.}\ }\textbf {\bibinfo {volume} {116}},\ \bibinfo {pages}
  {033002} (\bibinfo {year} {2016})}\BibitemShut {NoStop}%
\bibitem [{\citenamefont {Affolter}\ \emph {et~al.}(2020)\citenamefont
  {Affolter}, \citenamefont {Gilmore}, \citenamefont {Jordan},\ and\
  \citenamefont {Bollinger}}]{Affolter2020}%
  \BibitemOpen
  \bibfield  {author} {\bibinfo {author} {\bibfnamefont {M.}~\bibnamefont
  {Affolter}}, \bibinfo {author} {\bibfnamefont {K.~A.}\ \bibnamefont
  {Gilmore}}, \bibinfo {author} {\bibfnamefont {J.~E.}\ \bibnamefont {Jordan}},
  \ and\ \bibinfo {author} {\bibfnamefont {J.~J.}\ \bibnamefont {Bollinger}},\
  }\href {\doibase 10.1103/PhysRevA.102.052609} {\bibfield  {journal} {\bibinfo
   {journal} {Phys. Rev. A}\ }\textbf {\bibinfo {volume} {102}},\ \bibinfo
  {pages} {052609} (\bibinfo {year} {2020})}\BibitemShut {NoStop}%
\bibitem [{\citenamefont {Vasquez}\ \emph {et~al.}(2023)\citenamefont
  {Vasquez}, \citenamefont {Mordini}, \citenamefont {Verni\`ere}, \citenamefont
  {Stadler}, \citenamefont {Malinowski}, \citenamefont {Zhang}, \citenamefont
  {Kienzler}, \citenamefont {Mehta},\ and\ \citenamefont {Home}}]{Vasquez2023}%
  \BibitemOpen
  \bibfield  {author} {\bibinfo {author} {\bibfnamefont {A.~R.}\ \bibnamefont
  {Vasquez}}, \bibinfo {author} {\bibfnamefont {C.}~\bibnamefont {Mordini}},
  \bibinfo {author} {\bibfnamefont {C.}~\bibnamefont {Verni\`ere}}, \bibinfo
  {author} {\bibfnamefont {M.}~\bibnamefont {Stadler}}, \bibinfo {author}
  {\bibfnamefont {M.}~\bibnamefont {Malinowski}}, \bibinfo {author}
  {\bibfnamefont {C.}~\bibnamefont {Zhang}}, \bibinfo {author} {\bibfnamefont
  {D.}~\bibnamefont {Kienzler}}, \bibinfo {author} {\bibfnamefont {K.~K.}\
  \bibnamefont {Mehta}}, \ and\ \bibinfo {author} {\bibfnamefont {J.~P.}\
  \bibnamefont {Home}},\ }\href {\doibase 10.1103/PhysRevLett.130.133201}
  {\bibfield  {journal} {\bibinfo  {journal} {Phys. Rev. Lett.}\ }\textbf
  {\bibinfo {volume} {130}},\ \bibinfo {pages} {133201} (\bibinfo {year}
  {2023})}\BibitemShut {NoStop}%
\bibitem [{\citenamefont {Saner}\ \emph {et~al.}(2023)\citenamefont {Saner},
  \citenamefont {Băzăvan}, \citenamefont {Minder}, \citenamefont {Drmota},
  \citenamefont {Webb}, \citenamefont {Araneda}, \citenamefont {Srinivas},
  \citenamefont {Lucas},\ and\ \citenamefont {Ballance}}]{Saner2023}%
  \BibitemOpen
  \bibfield  {author} {\bibinfo {author} {\bibfnamefont {S.}~\bibnamefont
  {Saner}}, \bibinfo {author} {\bibfnamefont {O.}~\bibnamefont {Băzăvan}},
  \bibinfo {author} {\bibfnamefont {M.}~\bibnamefont {Minder}}, \bibinfo
  {author} {\bibfnamefont {P.}~\bibnamefont {Drmota}}, \bibinfo {author}
  {\bibfnamefont {D.~J.}\ \bibnamefont {Webb}}, \bibinfo {author}
  {\bibfnamefont {G.}~\bibnamefont {Araneda}}, \bibinfo {author} {\bibfnamefont
  {R.}~\bibnamefont {Srinivas}}, \bibinfo {author} {\bibfnamefont {D.~M.}\
  \bibnamefont {Lucas}}, \ and\ \bibinfo {author} {\bibfnamefont {C.~J.}\
  \bibnamefont {Ballance}},\ }\href {\doibase arxiv.2305.03450} {\  (\bibinfo
  {year} {2023}),\ arxiv.2305.03450}\BibitemShut {NoStop}%
\bibitem [{\citenamefont {Ospelkaus}\ \emph {et~al.}(2011)\citenamefont
  {Ospelkaus}, \citenamefont {Warring}, \citenamefont {Colombe}, \citenamefont
  {Brown}, \citenamefont {Amini}, \citenamefont {Leibfried},\ and\
  \citenamefont {Wineland}}]{Ospelkaus2011}%
  \BibitemOpen
  \bibfield  {author} {\bibinfo {author} {\bibfnamefont {C.}~\bibnamefont
  {Ospelkaus}}, \bibinfo {author} {\bibfnamefont {U.}~\bibnamefont {Warring}},
  \bibinfo {author} {\bibfnamefont {Y.}~\bibnamefont {Colombe}}, \bibinfo
  {author} {\bibfnamefont {K.~R.}\ \bibnamefont {Brown}}, \bibinfo {author}
  {\bibfnamefont {J.~M.}\ \bibnamefont {Amini}}, \bibinfo {author}
  {\bibfnamefont {D.}~\bibnamefont {Leibfried}}, \ and\ \bibinfo {author}
  {\bibfnamefont {D.~J.}\ \bibnamefont {Wineland}},\ }\href {\doibase
  10.1038/nature10290} {\bibfield  {journal} {\bibinfo  {journal} {Nature}\
  }\textbf {\bibinfo {volume} {476}},\ \bibinfo {pages} {181} (\bibinfo {year}
  {2011})}\BibitemShut {NoStop}%
\bibitem [{\citenamefont {Srinivas}\ \emph {et~al.}(2021)\citenamefont
  {Srinivas}, \citenamefont {Burd}, \citenamefont {Knaack}, \citenamefont
  {Sutherland}, \citenamefont {Kwiatkowski}, \citenamefont {Glancy},
  \citenamefont {Knill}, \citenamefont {Wineland}, \citenamefont {Leibfried},
  \citenamefont {Wilson}, \citenamefont {Allcock},\ and\ \citenamefont
  {Slichter}}]{Srinivas2021}%
  \BibitemOpen
  \bibfield  {author} {\bibinfo {author} {\bibfnamefont {R.}~\bibnamefont
  {Srinivas}}, \bibinfo {author} {\bibfnamefont {S.~C.}\ \bibnamefont {Burd}},
  \bibinfo {author} {\bibfnamefont {H.~M.}\ \bibnamefont {Knaack}}, \bibinfo
  {author} {\bibfnamefont {R.~T.}\ \bibnamefont {Sutherland}}, \bibinfo
  {author} {\bibfnamefont {A.}~\bibnamefont {Kwiatkowski}}, \bibinfo {author}
  {\bibfnamefont {S.}~\bibnamefont {Glancy}}, \bibinfo {author} {\bibfnamefont
  {E.}~\bibnamefont {Knill}}, \bibinfo {author} {\bibfnamefont {D.~J.}\
  \bibnamefont {Wineland}}, \bibinfo {author} {\bibfnamefont {D.}~\bibnamefont
  {Leibfried}}, \bibinfo {author} {\bibfnamefont {A.~C.}\ \bibnamefont
  {Wilson}}, \bibinfo {author} {\bibfnamefont {D.~T.~C.}\ \bibnamefont
  {Allcock}}, \ and\ \bibinfo {author} {\bibfnamefont {D.~H.}\ \bibnamefont
  {Slichter}},\ }\href {\doibase 10.1038/s41586-021-03809-4} {\bibfield
  {journal} {\bibinfo  {journal} {Nature}\ }\textbf {\bibinfo {volume} {597}},\
  \bibinfo {pages} {209} (\bibinfo {year} {2021})}\BibitemShut {NoStop}%
\bibitem [{Not()}]{Note1}%
  \BibitemOpen
  \href@noop {} {}\bibinfo {note} {See Supplemental Material for details of
  employed methods}\BibitemShut {NoStop}%
\bibitem [{\citenamefont {Wittemer}\ \emph {et~al.}(2019)\citenamefont
  {Wittemer}, \citenamefont {Hakelberg}, \citenamefont {Kiefer}, \citenamefont
  {Schr{\"{o}}der}, \citenamefont {Fey}, \citenamefont {Sch{\"{u}}tzhold},
  \citenamefont {Warring},\ and\ \citenamefont {Schaetz}}]{Wittemer2019}%
  \BibitemOpen
  \bibfield  {author} {\bibinfo {author} {\bibfnamefont {M.}~\bibnamefont
  {Wittemer}}, \bibinfo {author} {\bibfnamefont {F.}~\bibnamefont {Hakelberg}},
  \bibinfo {author} {\bibfnamefont {P.}~\bibnamefont {Kiefer}}, \bibinfo
  {author} {\bibfnamefont {J.~P.}\ \bibnamefont {Schr{\"{o}}der}}, \bibinfo
  {author} {\bibfnamefont {C.}~\bibnamefont {Fey}}, \bibinfo {author}
  {\bibfnamefont {R.}~\bibnamefont {Sch{\"{u}}tzhold}}, \bibinfo {author}
  {\bibfnamefont {U.}~\bibnamefont {Warring}}, \ and\ \bibinfo {author}
  {\bibfnamefont {T.}~\bibnamefont {Schaetz}},\ }\href
  {https://doi.org/10.1103/PhysRevLett.123.180502} {\bibfield  {journal}
  {\bibinfo  {journal} {Phys. Rev. Lett.}\ }\textbf {\bibinfo {volume} {123}},\
  \bibinfo {pages} {180502} (\bibinfo {year} {2019})}\BibitemShut {NoStop}%
\bibitem [{\citenamefont {Wittemer}\ \emph {et~al.}(2020)\citenamefont
  {Wittemer}, \citenamefont {Schr{\"o}der}, \citenamefont {Hakelberg},
  \citenamefont {Kiefer}, \citenamefont {Fey}, \citenamefont {Schuetzhold},
  \citenamefont {Warring},\ and\ \citenamefont {Schaetz}}]{Wittemer2020}%
  \BibitemOpen
  \bibfield  {author} {\bibinfo {author} {\bibfnamefont {M.}~\bibnamefont
  {Wittemer}}, \bibinfo {author} {\bibfnamefont {J.-P.}\ \bibnamefont
  {Schr{\"o}der}}, \bibinfo {author} {\bibfnamefont {F.}~\bibnamefont
  {Hakelberg}}, \bibinfo {author} {\bibfnamefont {P.}~\bibnamefont {Kiefer}},
  \bibinfo {author} {\bibfnamefont {C.}~\bibnamefont {Fey}}, \bibinfo {author}
  {\bibfnamefont {R.}~\bibnamefont {Schuetzhold}}, \bibinfo {author}
  {\bibfnamefont {U.}~\bibnamefont {Warring}}, \ and\ \bibinfo {author}
  {\bibfnamefont {T.}~\bibnamefont {Schaetz}},\ }\href {\doibase
  10.1098/rsta.2019.0230} {\bibfield  {journal} {\bibinfo  {journal}
  {Philosophical Transactions of the Royal Society A: Mathematical, Physical
  and Engineering Sciences}\ }\textbf {\bibinfo {volume} {378}},\ \bibinfo
  {pages} {20190230} (\bibinfo {year} {2020})}\BibitemShut {NoStop}%
\end{thebibliography}

%

\end{document}


\title{Supplemental Material for \\Phase-Stable Traveling Waves Stroboscopically Matched for Super-Resolved Observation of Trapped-Ion Dynamics}

\author{Florian Hasse}
\author{Deviprasath Palani}
\author{Robin Thomm}
\author{Ulrich Warring}
\author{Tobias Schaetz}
\affiliation{University of Freiburg, Institut of Physics, Hermann-Herder-Strasse 3, Freiburg 79104, Germany}

\begin{abstract}
In this Supplemental Material, we provide more insights into our experimental setup, sequences, parameter choices, and the pertinent interaction Hamiltonian. Furthermore, we present raw data and elaborate on our data analysis methods.
\end{abstract}
\date{\today}

\maketitle 

\section{Experimental Setup}
Our trap, housed in an ultra-high vacuum chamber, maintains a residual gas pressure below $10^{-8}$~Pa. 
We stabilize the lab environment's temperature to within $\pm0.3$~K to mitigate unwanted thermal variations. 
For more details about our specific experimental methods and apparatus, we refer to\,\cite{friedenauer_high_2006, clos_time-resolved_2016, Wittemer2019a}, while a general introduction to trapped-ion techniques can be found in Reference\,\cite{leibfried_quantum_2003}. 
Here, we choose the stretched states of the ground state $^2$S$_{1/2}$ manifold as our qubit states, i.e., $\ket{F=3, m_F = 3} = \ket{\downarrow}$ and $\ket{F=2, m_F = 2} =\ket{\uparrow}$, where $F$ is the total angular momentum and $m_F$ is the projection of the angular momentum along the magnetic field axis.
Notably, the designated manipulation periods within each sequence are shorter than any potential spin or motional dephasing durations, and from pure MW-controlled Ramsey measurements, we assess a coherence time of $\tau=70(1)\us$ for equal spin-superposition states, constrained by magnetic field fluctuations.

\begin{figure}
\begin{frame}{}
	\includegraphics[width=\textwidth/2]{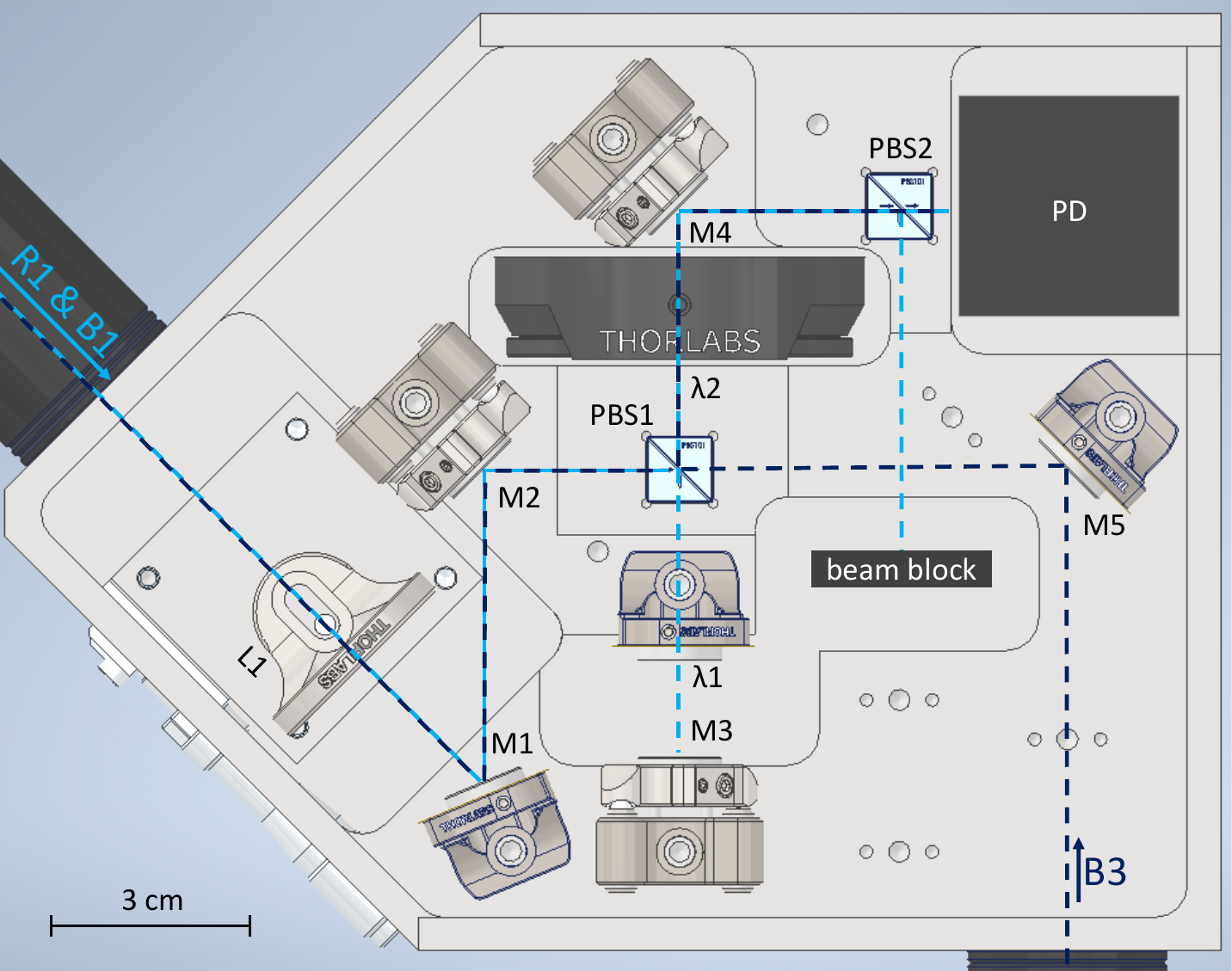}
\end{frame}
\caption{\label{fig-S1}
Our phase-stabilization setup. The optical components are arranged within an aluminum monolithic block. The beams R1 and B1 are overlapped, aligned with the ion, and directed towards the PD, facilitating phase-referenced CC control beams. In addition, the majority of the power of beam B3 is aligned anti-parallel to R1 near the ion, and a smaller fraction is overlapped with R1 on the PD, enabling phase-referenced AC control beams. The system allows for real-time switching between stabilized CC and AC beams.
}
\end{figure}

All relevant control equipment up to the GHz regime is actively phase-locked to a 10-MHz master clock. 
Our Global-Navigation-Satellite-System (GNSS) disciplined time and frequency reference comes with a referenced internal time base that has phase noise of less than -125\,dBc/Hz at 10\,Hz offset. 
To establish a phase-coherent link between the MW field and the TPSR beams, we undertake several measures: 
We utilize the same Direct Digital Synthesizer (DDS, phase referenced to the master clock) to generate the MW signal and control the AOMs of the TPSR beams, ensuring passive phase coherence. 
Additionally, we shield the laser beams with tubes from air disturbances. 
For active stabilization, we employ our phase-referencing setup, detailed in the main text, with a sketch of the mechanical set-up displayed in Fig.\,\ref{fig-S1}.

To actively combine AC and MW, the photodiode (PD) is illuminated with approximately 1 mW from R1 and approximately 0.2 mW from B3, yielding a heterodyne signal at 1.8\,GHz of approximately -60\,dBm, with the noise floor at approximately -100\,dBm. 
This signal undergoes filtering and amplification to approximately +15\,dBm, with a noise floor at approximately -25 dBm, before homodyning with the MW signal. 
A resulting error signal is employed through a servo lock to alter the AOM phase of one of the TPSR beams, ensuring short- to midterm phase coherence between AC and MW, a feature unattainable with solely passive stabilization -- see stability estimates below.
Similar power levels and procedures are used for CC beam configuration, cp. Fig.\,\ref{fig-S1}

\subsection{Stability Estimation of TPSR Phase}
We use the following Ramsey-type experimental sequences of pure or hybrids of MW and TPSR control fields to estimate phase coherence:
    \begin{itemize}
        \item Optional lock-up (TPSR-to-MW phase referencing) pulse ($\simeq100 \us$).
        \item Initialization in the spin state $\ketdown$ and close to the 3D (motional) vacuum state: Doppler cooling ($\simeq1530 \us$), and 3D sideband cooling ($\simeq 3-8 \ms)$ using AC configuration.
        \item First \pihalfpulse\ with MW ($\simeq 5 \us$) or TPSR beams ($\simeq2-3\us$): synchronizing the spin phase to the MW or TPSR control field with fixed phase.
        \item Variable waiting duration (0.01 to 100\,µs): for most experiments set to minimal duration, while varied to infer coherence durations.
        \item Second \pihalfpulse\ with MW or TPSR beams: analyze coherence, with variable phase $\phi$ or with fixed phase $\phiACnot$.
        \item Detection ($\simeq30\us$): detecting fluorescence photons are detected via the photomultiplier tube -- histograms are used to infer $\Pdown$.
    \end{itemize}
\begin{table}
    \centering
    \caption{Phase stability on short- ($2\s$), mid- ($40\s$), and longterm ($200\s$) scales for relevant coherent coupling field combinations.}
    \label{tab:stab}
    \begin{tabular}{lccc}
        \hline
        \hline
        & \multicolumn{3}{c}{Phase variances ($\deg$) for} \\
        & 2$\s$ & 40$\s$ & 200$\s$ \\
        \hline
        MW MW &  3.8(2)  &  1.17(6)  &  0.63(5) \\
        CC CC &  4.5(2)  &  1.15(6)  &  0.51(4) \\
        AC AC &  5.5(2)  &  1.93(9)  &  5.5(5) \\
        MW CC &  7.2(3)  &  1.49(8)  &  0.89(9) \\
        MW AC &  11.8(5)   &  22(1)  &  37(3) \\
        \hline
        \hline
    \end{tabular}
\end{table}
Refer to Table\,\ref{tab:stab} for pertinent benchmark figures regarding short-, mid-, and longterm stabilities. 
It is noteworthy that in all instances, we deduce a consistent coherence duration of approximately $70\us$, limited by magnetic field fluctuations, and we can prolong these coherence durations by factors of 10 to 20 by standard strategies, i.e., spin-echos or line-triggering.

\subsection{Static displacements -- scanning probe of the TPSR phase fronts}
In this section, we elaborate on the methodologies described around Fig.\,2(b) in the main text. 
We apply the Ramsey sequence with a combination of AC beams and a fixed phase $\varphi_0$. 
During initialization, we use dedicated shim fields and six control electrodes to displace the ion within the \xzplane, sampling at 670 positions using an adaptive technique\,\cite{Nijholt2019} and recording $\Pdown$. 
Interleaved reference experiments monitor residual TPSR phase drifts at the origin ($x_0$, $z_0$).
To realize a square scan area for Fig.\,2(b), the vertices of the sampling regions are pre-determined, the shim configurations are pre-aligned with $x$ and $z$ axes, and the applied shim amplitudes are calibrated for nm-precise displacements. 
The three-dimensional data are then analyzed using a variable cosine curve, adjusting amplitude, period, phase, and rotation in the \xzplane.
Fit results are given in the main text and align with our geometric expectations.

\section{Descriptions of parameters and the trapped ion Hamiltonian}
Our spin-motion system coupled via the TPSR polarization light pattern is described by the interaction Hamiltonian\,\cite{clos_time-resolved_2016}: 
\begin{equation}
\begin{aligned}
H_{\text{TI}} &= \hbar \omega_z \sigma_z/2 + \hbar \omega_m a^\dagger a \\
&\quad +  \hbar\Omega/2\,[C(\eta, a, a^\dagger)^\dagger \sigma_- + C(\eta, a, a^\dagger) \sigma_+ ].
\end{aligned}
\label{eq_HTI}
\end{equation}
The first term describes the spin, the second term represents the mode, and the third is the effective interaction engineered via our light fields. 
The parameters are explained as follows: $\hbar$ is Planck's constant, $\omega_z$ corresponds to the effective (dressed state) spin frequency and the motional mode is oscillating at $\omega_m$. 
The Rabi frequency $\Omega$, can be controlled by the intensity and detuning of our TPSR beams.
The annihilation and creation operators for the modes are expressed as $a$ and $a^\dagger$, while $\sigma_-$ and $\sigma_+$ represent the spin system lowering and raising operators. 
The Pauli $z$-operator is symbolized as $\sigma_z$. 
The Lamb-Dicke parameter is for our experimental parameters and the AC beam combination $\eta = \eta^{\text{AC}}_{\text{LF}} \simeq 0.4$, cp. Table\,\ref{tab01}.
We introduce a generic coupling operator, $C(\eta, a, a^\dagger)$, defined as $C(\eta, a, a^\dagger) = \exp[\mathrm{i} \eta (a^\dagger + a)]$.
\begin{table}
    \centering
    \caption{Properties of our coherent coupling fields: Frequencies $\omega$, Rabi rates $\Omega$, \effwl s $\leff$, and corresponding \LambDickeParameter s $\eta$ for all three motional modes.}
    \begin{tabular}{ccccc}
       \hline
       \hline
        & $\omega/(2\pi)$ & $\Omega/(2\pi)$& $\leff$ & $\LDLF$, $\LDMF$, $\LDHF$\\
        &in GHz&in MHz& & \\
       \hline
        MW & 1.8 & 0.1& $10\cm$ & $0.00, 0.00, 0.00$\\
        CC & 1.8 & 0.5& $\gtrsim80\um$ & $0.00, 0.00, 0.00$\\
        AC & 1.8 & 0.3& $140\nm$ & $0.40$, $0.23$, $0.18$\\
        \hline
        \hline
    \end{tabular}
    \label{tab01}
\end{table}
We numerically analyze the relevant dynamics of Eq.\,\ref{eq_HTI} via evaluations with QuTiP\,\cite{qutip2013}.

\section{Dynamic displacements -- Tracing motion in phase space}
Here, we extend our descriptions of the main text near Fig.\,3 and\,4: 
After the initialization, we use $D(\alpha)$ to coherently displace the ion along the LF mode.
The analysis pulse is applied stroboscopically with fixed timing and phase shift parameters of each pulse of the train. 
The total duration amounts to $t_{\pihalf,\text{S}}=\text{N}_{\text{S}}\cdot2\pi/\omegaLF\simeq30\cdot770\ns=23.1\us$, where the number of pulses $\text{N}_{\text{S}}$ is needed to yield a $\pi/2$ rotation with variable $\varphi$.
In the experiments, we optimize a $\pi/2$ pulse train using a for-loop that activates the AOMs of the two TPSR beams for a duration of $\delta t$. 
For each pulse of the train, we iteratively re-adjust the phase of the DDS controlling the AOMs, incorporating calculated delays.
Overall, timings and phases are optimized to effectively achieve a $\pi/2$ pulse using the experimental sequence with $\alpha = 0$.

\begin{figure}
\begin{frame}{}
	\includegraphics{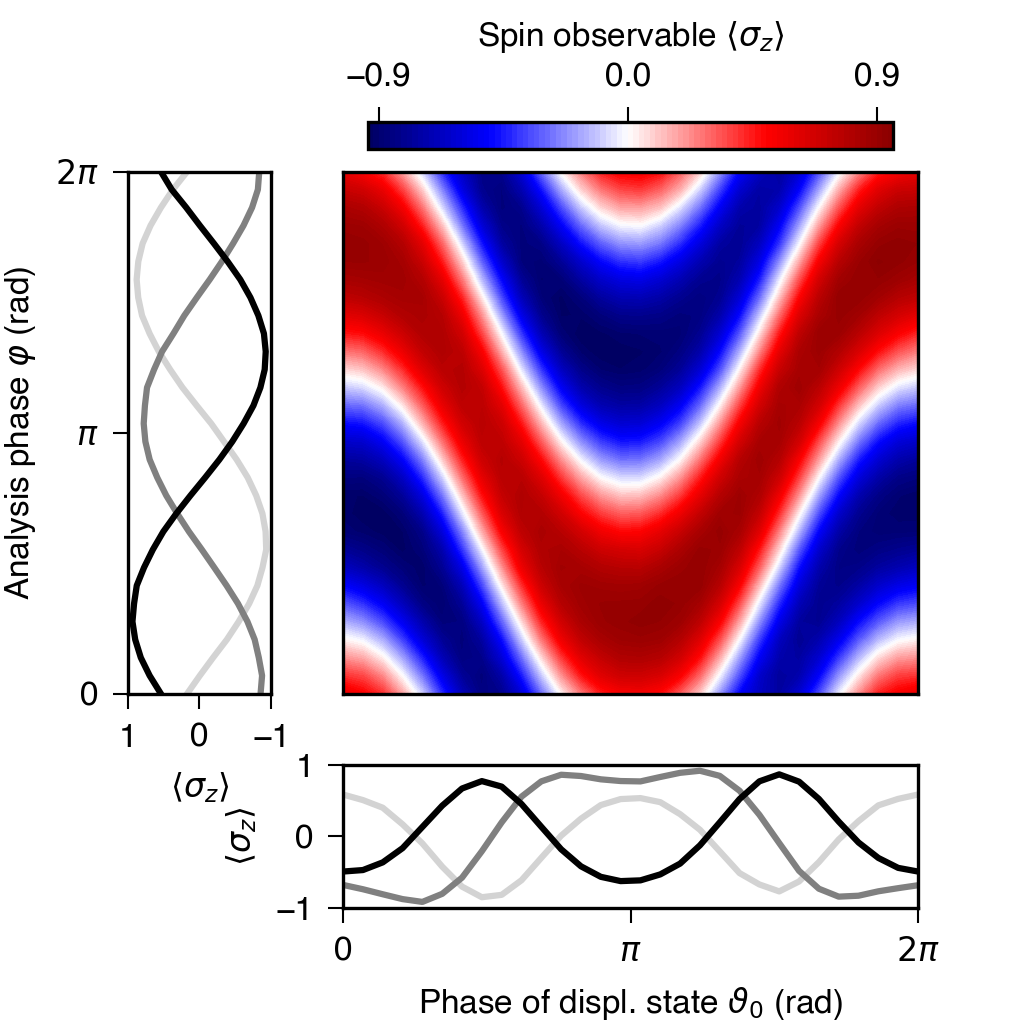}
    \includegraphics{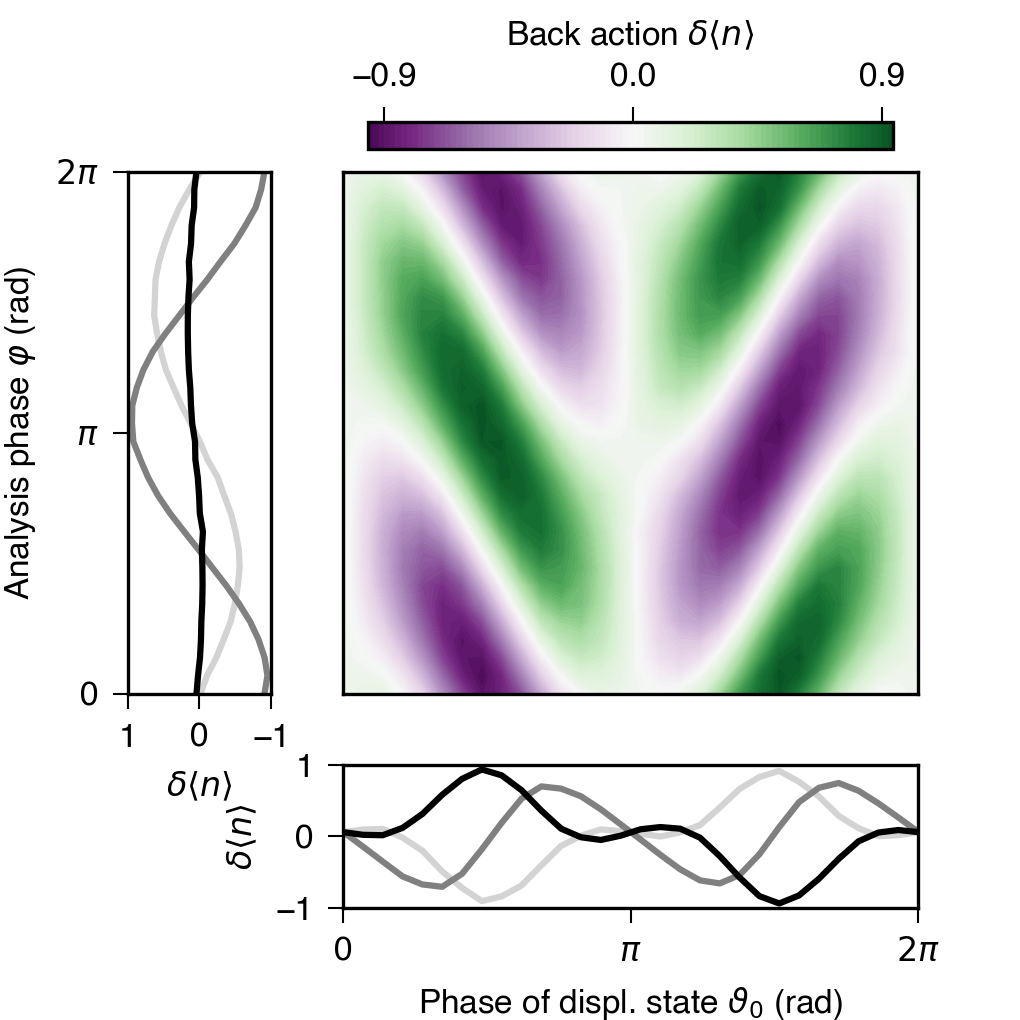}
\end{frame}
\caption{\label{fig-S2}
Illustration of the stroboscopic measurement principle, encoding position and momentum observables in the spin observable.
We perform numerical simulations to illustrate the principle of the experimental sequence, cp. Fig.\,3 of the main text.  
We show the detected spin observable $\langle\sigma_z\rangle$ and the amount of back-action $\delta \langle n \rangle$ (900 and 600 samples, linearly interpolated) as a function of $\varphi$ and $\vartheta_0$ for fixed displacement amplitude $|\alpha| = 3$, corresponding to coherent excitation of $\langle n \rangle_{\text{coh}} = 9$. 
Cuts along $\vartheta_0 = \{\pi/4, \pi/2, \pi\}$ and $\varphi = \{\pi/4, \pi/2, \pi\}$ highlight the underlying effects in light gray, gray, and black, respectively. 
}
\end{figure}

We quantify the encoding principle via numerical simulations and illustrate exemplary results in Fig.\,\ref{fig-S2}.
Here, we explicitly implement a time-dependent Rabi rate $\Omega = \Omega(t)$ matched to our experimental, stroboscopic parameters, see explanations above. 
We depict spin expectation values, $\langle \sigma_z \rangle$, as a function of analysis phase $\varphi$ and displaced state $\alpha$ parameters.
In addition, we evaluate the \emph{back action} of our stroboscopic coupling on the initial displaced state by the change in average motional quanta $\delta \langle n \rangle = \langle n \rangle_{\text{fin}} - \langle n \rangle_{\text{ini}}$. 
The initial $\langle n \rangle_{\text{ini}} \simeq \alpha^2$, while the final state is a non-trivial, but near-coherent, state with average quanta $\langle n \rangle_{\text{fin}}$.

\begin{figure}
	\includegraphics{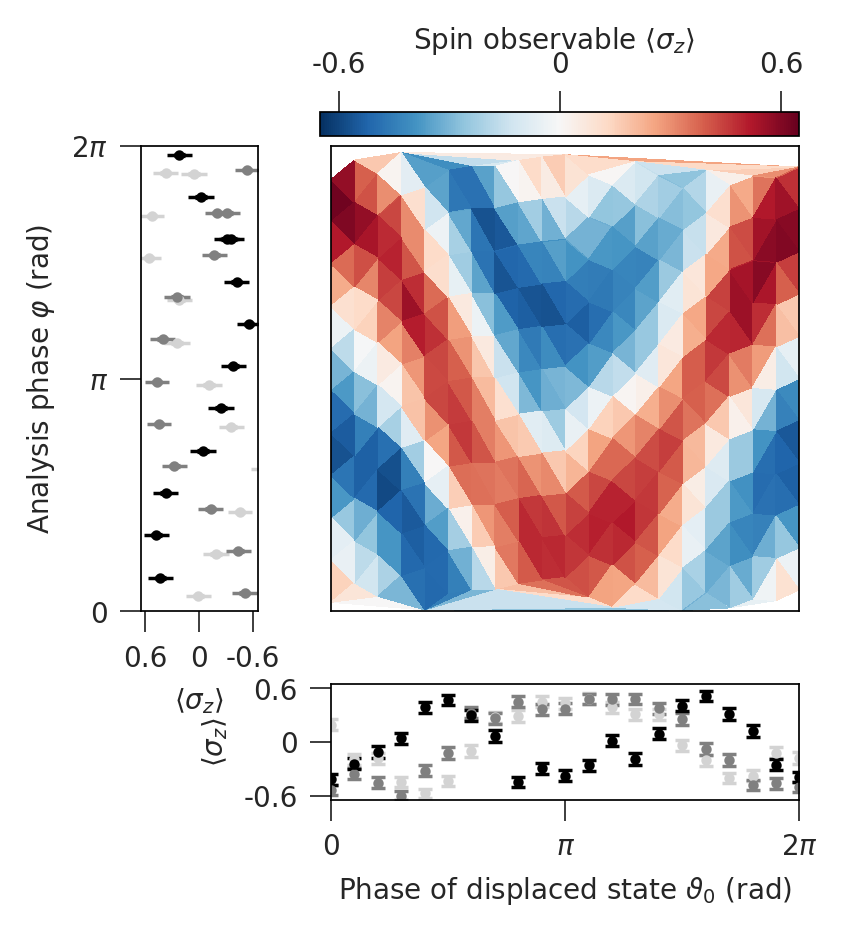}
\caption{\label{fig-S3_data}
Experimental raw data taken with the stroboscopic sequence, cp. Fig\,3(a). 
For fixed $\udispl=3.4(1)$, and variable $\varphi$ and $\vartheta_0$ (231 samples).
In addition, we show cuts along  $\vartheta_0 = \{\pi/4, \pi/2, \pi\}$ and $\varphi = \{\pi/4, \pi/2, \pi\}$ light gray, gray, and black, respectively.
This dataset can be compared qualitatively to numerical results shown in Fig.\,S2.
}
\end{figure}

To trace the ion's motional wave packet evolution in phase space, we require several preparatory steps and periodic recalibrations throughout the measurement series, with a focus on specialized procedures for stroboscopic measurement. 
Regular recalibrations involve, e.g., the qubit frequency, mode frequency, sideband cooling parameters, and initial motional quanta, alongside more basic measurements such as micromotion minimization and laser beam alignment.
In the measurement series, we sequentially interleave each experimental realization (approx. every 10 ms) with a phase-reference sequence with $\alpha = 0$, to re-adjust for residual AC phase variations.
Figure\,\ref{fig-S3_data} shows experimental data of $\langle \sigma_z \rangle$ for fixed $\udispl=3.4(1)$, and variable $\varphi$ and $\vartheta_0$.
Figures\,S2 and S3 can be compared qualitatively, numerical results represent ideal-case scenarios, ignoring relevant dephasing effects.

\begin{figure}
\begin{frame}{}
	\includegraphics{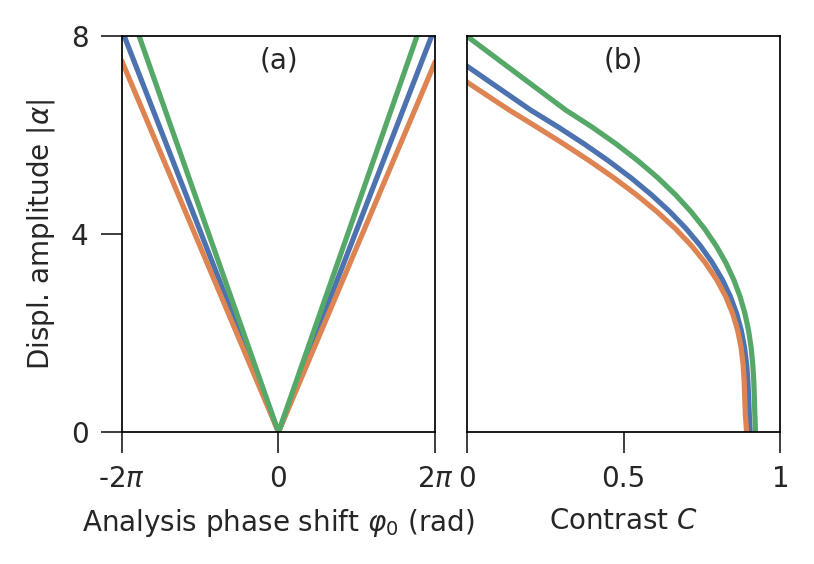}
\end{frame}
\caption{\label{fig-Calibration_UW} Numerical calibrations to convert analysis phase shifts $\varphi_0$ and contrast $\contrast$ into position (a) and momentum, (b), respectively for a single $^{25}$Mg$^+$ ion with initial thermal $n_\text{th}=0.15$.
To illustrate the effect of the uncertainty on the LF mode angle regarding the $z$ axis, we show calibrations for $0\,\deg$ (blue), $-5\,\deg$ (green), and $+5\,\deg$ (orange).
}
\end{figure}

In our analysis, we utilize numerical simulations to deduce necessary calibrations for estimating position and momentum observables from spin projection results, illustrated in Fig.\,\ref{fig-Calibration_UW}. 
Notably, all parameters, with uncertainties around a few percent, are independently measured. 
An independent calibration series of displacement amplitudes using red and blue sideband analysis\,\cite{leibfried_quantum_2003} is in agreement with assumed amplitudes, considering an LF mode angle orientation of $0(5) \deg$ regarding the $z$ axis. 
\section{Comments on Squeezed states}
\begin{figure}
\begin{frame}{}
	\includegraphics{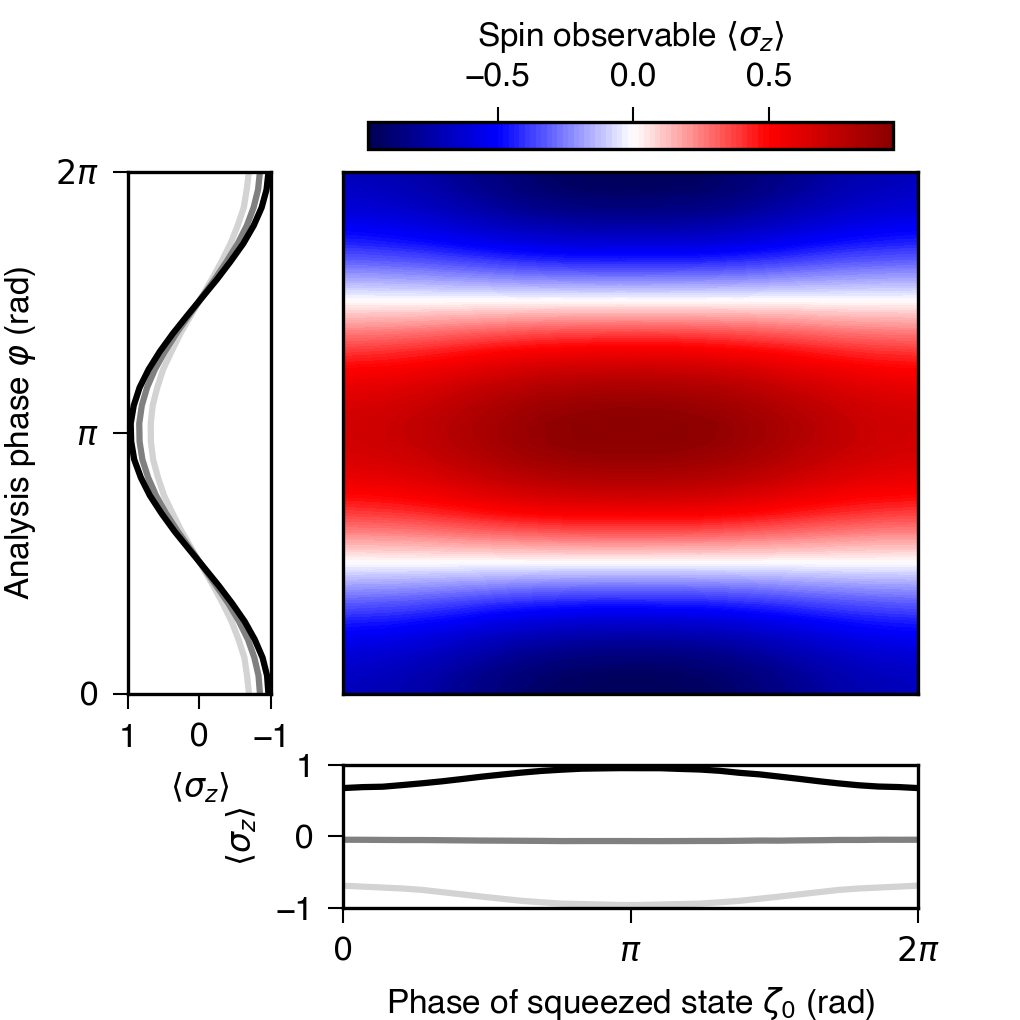}
    \includegraphics{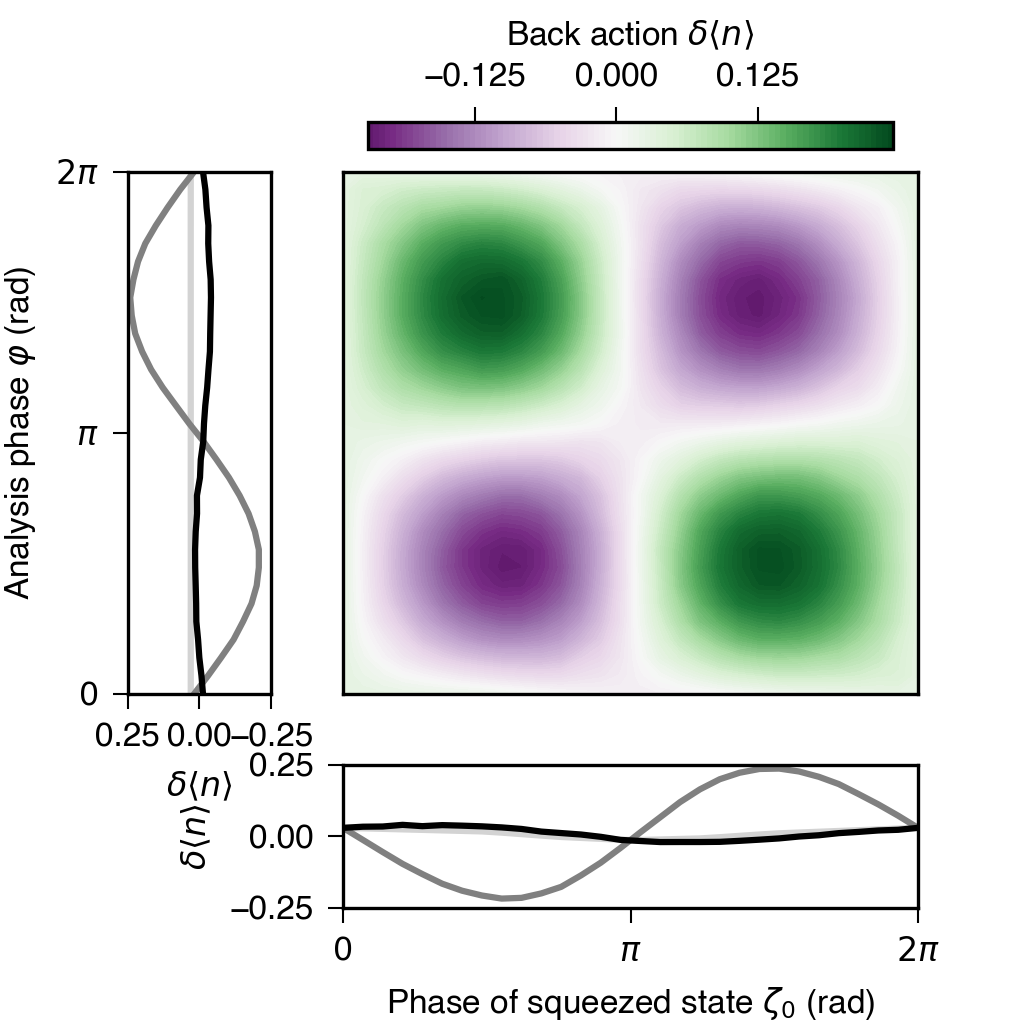}
\end{frame}
\caption{\label{fig-S4}
Illustration of the stroboscopic control and measurement principle for squeezed states.
We perform numerical simulations to illustrate the possibility of resolving and controlling squeezed state dynamics.  
We show the spin observable $\langle\sigma_z\rangle$ and the amount of back-action $\delta \langle n \rangle$ (600 and 300 samples, linearly interpolated) as a function of $\varphi$ and $\zeta_0$ for fixed squeezing amplitude $|\zeta| = 1$ (average motional quanta $\langle n \rangle_{\text{sq}} = 1$). 
Cuts along $\zeta_0 = \{0, \pi/2, \pi\}$ and $\varphi = \{0, \pi/2, \pi\}$ highlight the underlying effects in light gray, gray, and black, respectively. 
}
\end{figure}

In the future, the application of our stroboscopic traveling wave pattern to squeezed states becomes relevant. 
A squeezed vacuum state is obtained by applying the squeezing operator $S(\zeta)$ to the vacuum state $|0\rangle$. The squeezing operator is defined as:
\[S(\zeta) = \exp\left(\frac{1}{2}(\zeta^* a^2 - \zeta {a^\dagger}^2)\right),\]
where $a$ and $a^\dagger$ are the annihilation and creation operators, respectively. 
We define the parameter $\zeta = |\zeta| e^{i\zeta_0}$, where $|\zeta|$ is the squeezing amplitude and $\zeta_0$ is the squeezing phase. 
A squeezed vacuum state $|\zeta\rangle$ is then $|\zeta\rangle = S(\zeta)|0\rangle$. 
We apply a similar stroboscopic pulse train as above, but with $\Delta t = 2\cdot2\pi/\omegaLF$ in our numerical code to such states, and we give some illustrations of the principles in Fig.\,\ref{fig-S4}.
From these illustrations, we see, that only contrast variations can be used to infer the squeezed state dynamics: Squeezed positions $\zeta_0 = \pi$ yield higher contrast, and squeezed momenta $\zeta_0 = 0$ yield lower contrast than the vacuum states.
The amount of back-action can be tuned via the analysis and/or squeezed phase adjustments, sketching the foundation for advanced control schemes, while details need to be studied further in future work.


%